# Living Innovation Laboratory Model Design and Implementation

[A Book Draft]

By

Yuting Zheng

27 Jan 2016

# Abstract

Living Innovation Laboratory (LIL) is an open and recyclable way for multidisciplinary researchers to remote control resources and co-develop user centered projects. In the past few years, there were several papers about LIL published and trying to discuss and define the model and architecture of LIL. People all acknowledge about the three characteristics of LIL: user centered, co-creation, and context aware, which make it distinguished from test platform and other innovation approaches. Its existing model consists of five phases: initialization, preparation, formation, development, and evaluation.

However, it has some drawbacks. LIL relies on user requests, which usually results in incremental innovation, instead of disruptive innovation. Unlike incremental innovation improving the existing market, disruptive innovation can expand the markt and even create a new market. It requires us to discover and fulfill user needs, no matter users realize what they want or not. In addition, LIL co-creation team is a targeted group of users, developers and industry party, which may limits the creativity. Plus, as technology is chaning rapidly, instant omtext awareness may not be enough.

Therefore, a new generation of LIL emerges, that is LIL 2.0. Its characteristics include unobtrusive user involvement (UUI), massive co-creation (MCC), and predictable context aware (PCA). UUI helps to discover the hidden user needs. MCC makes the co-creation team more diverse. PCA makes the innovation proactive and forward-looking. Thus, LIL 2.0 has more chance to produce disruptive innovation in an effective and efficient way with lower business risk. Some advanced concepts, such as big data, crowd sourcing, crowd funding and crowd testing, can facilitate UUI, MCC and PCA and eventually help to build LIL 2.0.

Goal Net is a goal-oriented methodology to formularize a progress. In this thesis, Goal Net is adopted to subtract a detailed and systemic methodology for LIL. LIL Goal Net Model breaks the five phases of LIL into more detailed steps. Big data, crowd sourcing, crowd funding and crowd testing take place in suitable steps to realize UUI, MCC and PCA throughout the innovation process in LIL 2.0. It would become a guideline for any company or organization to develop a project in the form of an LIL 2.0 project.

To prove the feasibility of LIL Goal Net Model, it was applied to two real cases. One project is a Kinect game and the other one is an Internet product. They were both transformed to LIL 2.0 successfully, based on LIL goal net based methodology. The two projects were evaluated by phenomenography, which was a qualitative research method to study human experiences and their relations in hope of finding the better way to improve human experiences. Through phenomenographic study, the positive evaluation results showed that the new generation of LIL had more advantages in terms of effectiveness and efficiency.

In conclusion, this thesis did not only review the literature of LIL 1.0, but also studied the evolution from LIL 1.0 to 2.0. Then a detailed methodology for LIL was proposed and applied it to two real world projects. This thesis suggested some advanced concepts, such as big data and crowd sourcing, to be used in LIL 2.0, but did not look into the details of each concept. It might be the area of concern in our future works.

# Acknowledgements

I would like to take this opportunity to send out my thanks to Dr. Miao chunyan, my supervisor. Not only did she introduce me to the research field, she was always willing to discuss with me no matter how strange my questions are. I would like to thank her for her patience and efforts to show me the way of doing research, her careful review of every little piece I write and her valuable suggestions for my research directions.

Also I would like to extend my appreciation to my colleagues in LILY research centre. They spent a lot of effort on helping and supporting me to practice my methodology in their project. I also got generous help from my team members in Baidu during the product development process.

Special thanks go to Andrew Kusiak and Dr. Shen Zhiqi for their inspiring papers.

# Contents

# List of Figures

# List of Tables

# Nomenclature

LIL – Living Innovation Laboratory

UUI – Unobtrusive User Involvement

MCC – Massive Co-creation

PCA – Predictable Context Aware

LIL 1.0 – The traditional Living Innovation Laboratory

LIL 2.0 – The new Living Innovation Laboratory

KG – Knowledge Graph

CTR – Click-Through-Ratio

# Chapter 1:

# Introduction

This chapter introduced the background information of the thesis, including how the concept of Living Innovation Laboratory (i.e. LIL) emerged and the motivation of this thesis. Tow research questions were addressed here and would be further discussed in the following chapters. The key contribution of the thesis is to propose the new generation of LIL and a detailed methodology to build it based on goal net model.

## 1.1 Research Background

Nowadays, user market is changing so fast that technology innovation cannot follow on time. All over the world, eighty-five percents of the research efforts in the world are spent on products and services that eventually fail to apply to the real market. Meanwhile, the potential of certain products or services (such as mobile payment) is totally underestimated by the experts before, thus unexpected market opportunities are missed. In order to make a useful invention that can be beneficial to more people, we need to overcome the gap between what the developers are producing and what people actually need from the beginning of the project.

The concept of LIL first emerged in Europe. It introduced a new approach to stimulate user-driven innovations in order to better understand and exploit innovations. The key characteristics of LIL include user centered, co-creation, and context aware. Based on "user centered", LIL is able to create requirement-driven innovation, which more directly fulfills user requirements compared with theoretical data-driven innovation and passive technology-driven innovation. "Co-creation" makes LIL distinguished from test bed, field trial and other testing platforms which only involve users in certain sections instead of the whole innovation process. "Context aware" means that LIL should adapt to any contextual changes so that its products can keep pace with the times. Some previous papers about LIL highlighted the three characteristics and proposed some high-level frameworks to build an LIL. Generally, an LIL runs its business through five phases: initiation, preparation, formation, development, and evaluation. There are three parties participating in LIL, including users, developers, and industry party. All of them should be involved from the beginning phase until the end of innovation life cycle, which represents "user centered" and "co-creation".

## 1.2 Issue and Challenges

Although the traditional model of LIL which was first introduced in Europe years ago was a great leap forward, it might not be the most suitable one, especially for today. Today, more and more innovative concepts (such as Smart Phone and Internet Finance) are generated even beyond user requests. In the traditional model of LIL (i.e. LIL 1.0), innovation is driven by user feedbacks and co-created by a targeted group of users, developers and industry party. Even if user feedbacks were collected in the first place, it was still hard to imagine the concept (for example, smart phone) which users themselves did not realize that they need it. Only when the first smart phone (i.e. iPhone) was published to users, they were surprised that it was exactly what they wanted. LIL 1.0 can bring users what they request, but cannot bring them what beyond their imagination. Therefore, LIL 1.0 tends to result in incremental innovation, instead of disruptive innovation [1]. Incremental innovation refers to the continuous improvement of a product within the existing market, while disruptive innovation means an innovative product which can expand the boundary of the existing market or even create a new market. In addition, today's technology (such as Mobile Internet and Wearable Devices) innovates much faster than expected. "Context aware", which guides us to react once we realize the change, may be insufficient for today. In order to catch up the innovation speed and keep ahead in the market, companies had better take actions before the changes happen. Based on the challenges above, we realized that the traditional model of LIL should be refined.

Furthermore, as more and more enterprises and regions start to commit themselves to building LILs or join the network of LILs, the requirement for the basic principles and guidelines of LILs are emerging. The previous papers have introduced us the elements, construction and workflow of an LIL from a high-level perspective. However, there is no detailed framework or methodology about how to build a qualified LIL in a professional but easy way.

Hence, two research questions were derived from the issues mentioned above:

RQ 1: Can we refine the three characteristics of LIL? Can we define a new generation of LIL with the new characteristics?

RQ 2: Can we have a detailed methodology to build this new generation of LIL?

To answer these two questions, research works were conducted and summarized in this thesis. It contributed to a better understanding and successful use of LIL as an innovation approach.

## 1.3 Contribution

This thesis tried to outline what the traditional LIL (LIL 1.0) is, how to refine it into the new generation of LIL (LIL 2.0), and then introduced a new methodology to design and develop LIL 2.0 in a systemic way. All of the issues mentioned above were addressed and solved as follows:

- Issue 1: "user centered" in LIL 1.0 may result in incremental innovation.
    - ✓ Solution: "unobtrusive user involvement" tries to collect user data no matter they are aware or not, which may reveal comprehensive user demands in an unobtrusive way. For example, user behavior big data may indicate what users actually need, even though they do not realize and cannot feedback about that. Here, big data technology could be one of useful approaches to achieve "unobtrusive user involvement". By understanding what users actually need but cannot imagine themselves, people tend to create disruptive innovation and a new market beyond user's expectation [1].
- Issue 2: "co-creation" in LIL 1.0 within a targeted group of people may be too narrow.
    - ✓ Solution: "massive co-creation" means that a large number of users, development talents and industry parties are gathered to jointly develop a project, usually remotely [9]. Crowd sourcing, crowd funding and crowd testing are three ways to realize massive co-creation. Based on the more diverse sample and crowd talents, LIL becomes more effective and efficient.
- Issue 3: "context aware" in LIL 1.0 may fail to catch up the fast innovation pace today.

- ✓ Solution: "predictable context aware" leads people to make use of big data to predict the contextual change and prepare the corresponding strategies before the change comes. It would make LIL more adaptable and sustainable in the changing world.

■ Issue 4: the previously proposed methodology to build LIL is too high-level.

- ✓ Solution: LIL Goal Net Model was proposed in the thesis. Basically, Goal Net Model is a goal-oriented methodology to formularize a progress [30]. Now it is used to subtract the detailed methodology for LIL 2.0. It would become a guideline for any party or organization to develop a project in a living ecosystem and adapt to the fast changing world.

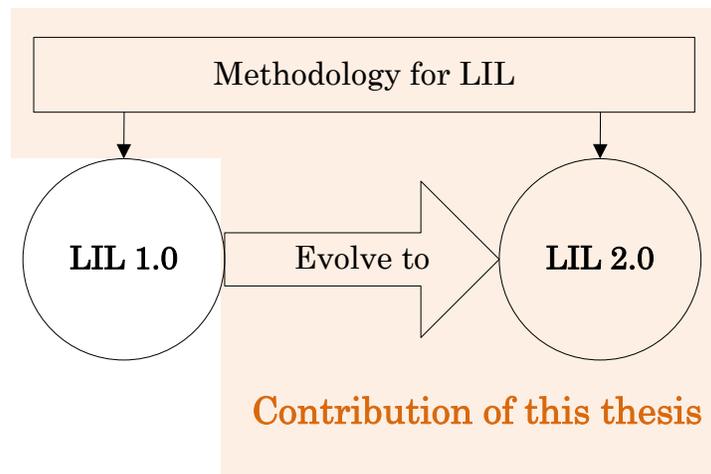

Figure 1.1: The contribution of this thesis

Figure 1.1 above illustrates our area of concern. Throughout a sequence of research works on the area of concern, the two research questions were answered as follows:

RQ 1: Can we refine the three characteristics of LIL? Can we define a new generation of LIL with the new characteristics?

Answer: Yes, a new generation of LIL (i.e. LIL 2.0) was derived from the research and analysis. It has three newly defined characteristics: unobtrusive user involvement, massive co-creation, and predictable context aware. LIL 2.0 is advanced than LIL 1.0, in terms of effectiveness, efficiency, and sustainability.

RQ 2: Can we have a detailed methodology to build this new generation of LIL?

Answer: Yes, the proposed methodology illustrated the high-level five phases in LIL can be broken down into detailed steps and what actions should be taken in each single step to build a qualified LIL 2.0. Advanced technologies, such as, big data, crowd sourcing, crowd funding, and crowd testing, should be used in suitable steps.

Finally, to demonstrate these solutions and answers were correct, two real world projects were implemented based on the concept of LIL 2.0 and the proposed LIL methodology. One project was a Kinect game. The survey among project team members showed the positive of LIL 2.0 and the proposed methodology. The other project was an Internet product, named "Knowledge Graph" in search engine. After applying LIL Goal Net Model, this project was changed to LIL 2.0 and achieved a better result measured by experiment data.

## 1.4 Organization of the Paper

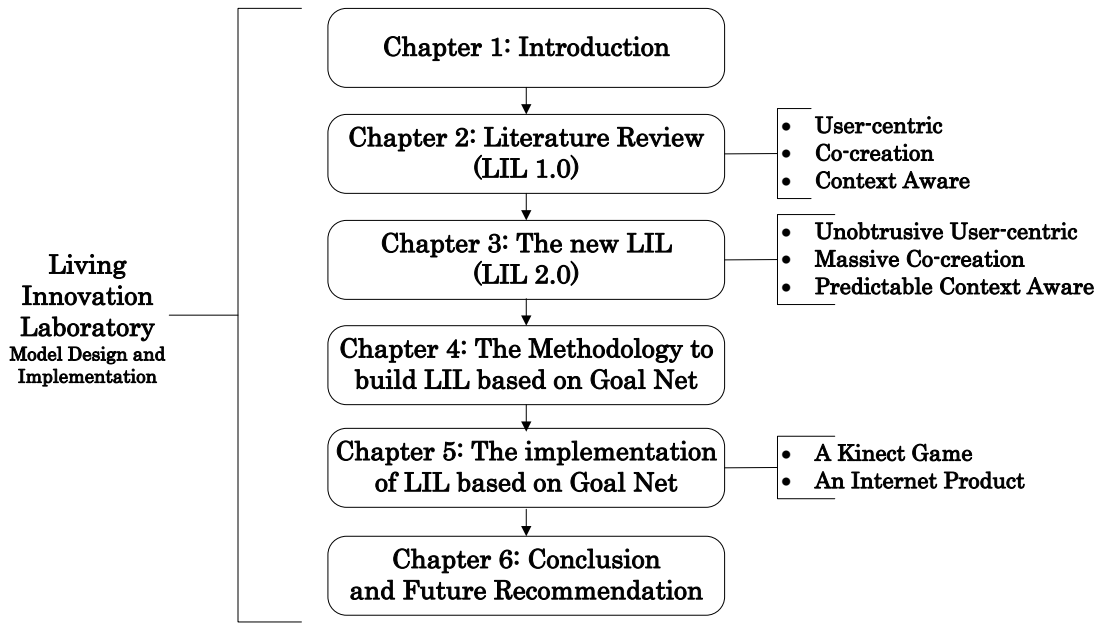

Figure 1.2: The organization of the thesis paper

The remainder of this report is organized as follows: Chapter 2 begins with a literature review of existing definition on LIL (LIL 1.0), including what it is and how it works, as well as Goal Net Model, including its theory and its use. The new generation of LIL (LIL 2.0) is described in Chapter 3. Next, a methodology for LIL 2.0 design will be introduced and illustrated in Chapter 4. Chapter 5 used an LIL project to show how to apply this methodology to real case and demonstrate how LIL 2.0 concept facilitated the innovation. Chapter 6 summarized the core contributions made in the present research, outline potential directions for future research and provides the major conclusions.

# Chapter 2:

# Literature Review

This chapter reviewed many definitions of LL from different perspectives. The three characteristics of LIL showed its differences from other research and development approach [7]. Through the typical model and high-level architecture of LIL, we could see how an LIL project works and how different players work together in the life cycle of an LIL project.

## 2.1 What Living Innovation Laboratory is

The concept of LIL was first introduced by Prof William Mitchell, in MIT, Boston. He described LIL as a user centered innovation approach for designing, developing and validating new technologies, products and services with end users in real life contexts [3].

Later, a group of European organizations established the first LIL in the world. They emphasize user contribution and participation in the innovation process. The first LIL project was about smart homes in the future. Users were invited to stay in the experimental home setting for several days or weeks. By observing user behaviors and collecting feedback from users, researchers better undertook what users expected and how to use emerging technologies to fulfill user requirements. Meanwhile, users were free to express and contributed their ideas to the project as a co-creator.

### 2.1.1 Definition of Living Innovation Laboratory

Living Innovation Laboratory (LIL) refers to a research concept about an open and recyclable way for multidisciplinary researchers to remote control resources and co-develop user centered projects.

Generally, there are three participants involved in LIL.

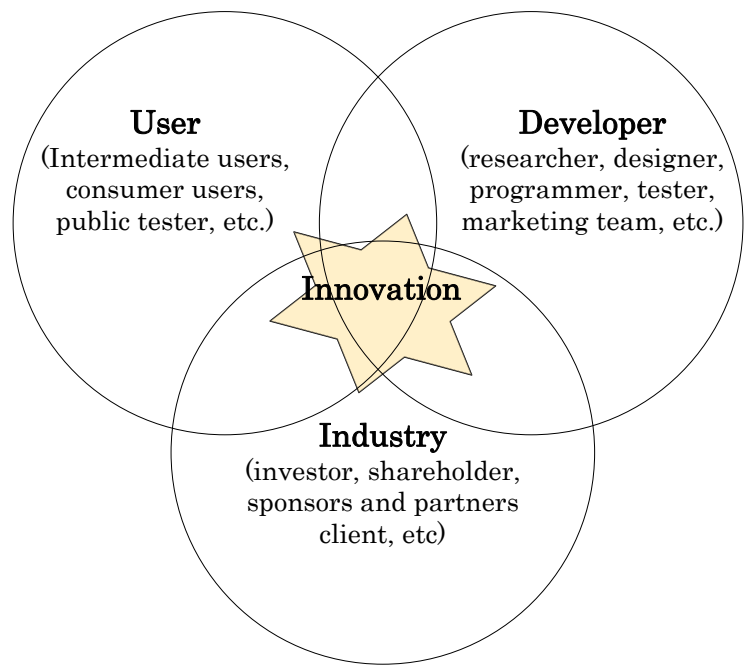

Figure 2.1: Three Parties involved in Living Innovation Laboratory

The implementation of LIL is based on the involvement of three parties: **user**, **developer** and **industry party** in the innovation process. They interact with each other to find out the problem and actively work together towards the solutions.

As we noticed from the elements and players in LIL, a bunch of intelligent **resources**, usually remote, are collected and shared in LIL. These "intelligent resources" do not only include technical resources (for example, knowledge, innovative ideas, technologies, infrastructures, and so on), but also human resources (for example, science communities, partnerships, business networks, users with their experiences, and so on). The project initiators are responsible for gathering those intelligent resources or organizing them if they are attracted due to their own interest. The "organizing" means the resources, once involved in the project, would become saved for reusing, open for sharing and studied for keep-going improving.

In a word, LIL creates a **collaborative** research environment for a human centric project with various resources (including knowledge and expertise) evolved. Hence, the project initiator may organize a multidisciplinary team (involving user, developer, and industry party) to co-develop a human centric product or service in a living ecosystem.

### 2.1.2 Difference from other testing platforms

There are six different kinds of user centered testing platform, including test bed, field trial, market pilots, societal pilots, and LIL [5]. By comparing and differentiating them, we may have a better understanding about the specialties and advantages of LIL:

1. **Test bed** aims to test something already developed. In contrast, LIL usually focuses on how to design a new product.

2. **Field trial** is to try different exiting things at the same time to find which one is the best, while LIL is used to inspire people's ideas and facilitate co-creation towards a new product.

3. The commercial maturity of what is tested is normally higher in the **societal and market pilots** compared to in LIL [3].

As we can see, LIL is more design focus and open-ended. The solutions or even problems are not defined clearly yet. In this open innovation context, mo re creative ideas may be inspired. Users in a LIL are not only considered as observed subjects only for testing purpose, but turn to be co-creators and even co-developers who contribute creative ideas, innovative concepts, development works and so on.

## 2.1.3 Difference from other innovation approaches

Before LIL was introduced, the most widely-used approaches towards innovation were data-driven innovation and technology-driven innovation. What are the differences between LIL and these two approaches?

**Data-driven innovation** usually happens in universities and some R&D labs. Researchers search all kinds of legacy materials in standard and digital libraries, in the quest of innovation. Some algorithms, such as data mining, test mining, and information analysis could be helpful for abstract previously unseen value in fusing data and discover new knowledge from various sources (Andrew Kusiak, 2007). For example, Einstein discovered the theory of relativity; Watson and Crick discovered the molecular structure of DNA. They revolutionized their fields and brought innovations that were unexpected, but nevertheless did not affect existing markets. In comparison, LIL aims to create an innovation that improves a product or service in an existing market in ways that users are expecting. To achieve it, LIL needs to make use of different methods including data-driven research and user-driven research.

**Technology-driven innovation** is to use the new technology to affect the existing market, but nevertheless does not meeting user expectations. The innovation process is initiated by a certain degree of technological breakthrough, such as a new mechanism or device, and followed by a series of developments. LIL is based on market-pull theory, instead of the technology-push approach. Market demands and user requirements are the source and driving force of innovation. Therefore, LIL involves users in the beginning stage and all the way until the final testing and evaluation stage. It uses the novel ideas and diverse resources to affect the existing market as user expected, but nevertheless does not involving new technology. For example, Mark Zuckerberg created Facebook by using the existing Internet technology, which does not only fulfill the user needs of social communication but also leads to an evolution of human social networking.

Actually, LIL introduced a new concept "**Requirement-driven innovation**", which means the innovation is driven by the requirements from user, community, market, economy and other parties. In either data-driven innovation or technology-driven innovation, the "engineer-as-the-king" model allows technical experts make the decisions for the customer. But in LIL, requirement-driven innovation realizes the "customer-as-the-king" model, which allows the customer to request and co-create new products or services rather than just accepting the offered ones. Furthermore, the user involvement in the innovation process may ensure highly reliable market evaluation, and reduce business risks, and thus save development costs [1].

## 2.2 The characteristics of Living Innovation Laboratory

The word of "living" in LIL makes it extremely different from any other innovation laboratory. There are three characteristics of LIL which just perfectly illustrate the meaning of "living".

### 2.2.1 User centered

In LIL, both of the collaborative manner of development and the openness of living innovation require "**user centered**". User centered approach makes researchers to hear the user feedback on existing products and then user demand or expectation for future products. Users do not only play an important role in product review but also the value-creation chain, throughout the whole process of product creation. The intention of involving users is to understand what people want, what kind of innovation is worth enough of continuous developing, and eventually can serve people for a long term. The context of a changing society makes any new technical product hardly match people's requirement forever. Any technical breakthrough aims to solve people's current problem, which demands researchers to include users as a part of development team and hear their voice all the time, just like developers. This is just the basic idea of user-centered method, which may help researchers to better define a product in the initial stage and often refine a solution in the subsequent steps. After all, all potential innovations emphasize how they serve users better, rather than how technically superior they can perform.

If a successful innovation should bring some value to users, a research life-cycle should be user centered, not technology centric any more. Thus, the active and interactive involvement of users in LIL must be far beyond traditional mass surveys or focus groups. In LIL, users (including intermediate users, consumer users, and public tester) are now expected to implement an innovation together with developers.

Anna Ståhlbröst (2008) categorized different degrees of user involvement into the three clusters [7]:

1. <u>Design *for* users</u>: The users are guided by the development team and following the steps to give feedbacks. This kind of **passive** participating usually occurs in the late stage of the development process, for the purpose of validating requirement specifications and testing prototypes. For exampe, game players are invited to testing alfa or beta version of a new game.

2. <u>Design *with* users</u>: The users participate in the whole process of development and join the design of the future prodcut by expressing their needs and expectations. The users **actively** take in charge of the context and evaluation activities, but the design and development activities are still controlled by the development team. For example, game players co-create the storyline of a massively multiplayer online roleplaying game.

3. <u>Design *by* users</u>: The users **drive** the project by contributing ideas and developing products or parts of products. The development team direct where to go, while the users decide how to go. For example, game players propose a game as IP (intellectual property) and help the game company to develop it.

As we can see, the degree of user involvement increases from 1 to 3. "Design *for* users" usually happens in testing platforms. LIL tries to reach all of the three degree, especially "Design *with* users" and "Design *by* users". To achieve it, LIL makes use of vairous user data collection methods, such as focus-group, one-to-one interviews, online or offline surveys, and work-shops, from the beginning of project all the way to the final testing stage.

## 2.2.2 Co-creation

In additional to involving users in the project design, development and testing processes, LIL also strives to facilitate the interaction between other relevant industry parties, such as, shareholders, sponsors, public and civil sectors and the society (Feurstein, Hesmer, Hribernik, Thoben, and Schumacher 2008).

LIL's open and multidisciplinary environment encourages people (including users, developers and industry parties) to collaborate in LIL system and eventually **co-create** a comprehensive project. At this point, the involvement of end-users and stakeholders both into the development process gives a new definition of co-creation methodology. The experiments conducted among user groups and the evaluation reports by sponsor groups do not only play as an observed feedback but also as a source of creation. Thus, the developers can take multi-contextual factors into their consideration, including the eventual usefulness of a product or service, the comments by user groups, and the satisfaction of sponsor groups. This kind of concurrent consideration may happen in either early or recycling stage of the produce life-cycle, and finally help developers to justify and improve the overall performance of a product or service. Overall, LIL provides a flexible co-creation way to complete a comprehensive innovation process.

Hence, LIL enables different groups to perform their distinctive functions and work together throughout the entire value chain, including fund suppliers, content generators, technology providers and the end-users.

### 2.2.3 Context aware

Professor William Mitchell, as the father of LIL, described the concept of LIL as "a user centered research methodology for sensing, prototyping, validating and refining complex solutions in multiple and evolving **real life contexts**" (Eriksson et al, 2005, p. 4). Real life contexts, including user preference, emerging technology, investment market, political issue, nature environment and so on, actually make the innovation lab "living".

Nowadays, user preference is changing any time, technology is also updated continuously, and the market is changing all the time. Sometimes, a product needs to be updated to the next version while the first version is just released, because the user preference has changed. Sometimes, the design of a product must be totally changed even before its release, because a new technology is coming out. Once the update cycle is stopped or becomes infrequent, the product may be out of date or lose the market. For example, since the first version of iPhone was released and became popular in 2009, the era of smart phone was coming. "Samsung", immediately changed their hand phone design from keyboard pattern to touch-screen pattern. "Blueberry", did not change their hand phone pattern according to the change of user market on time. Nowadays, Samsung becomes the second large hand phone producer. In contrast, Blueberry has totally lost its empire in hand-phone market.

So, whatever innovation topic an LIL is working on, it should be aware of the real life contexts and flexibly adapt itself to any change of the multi-contextual environment. In conclusion, the three characteristics of LIL are User-centered, Co-creation, and Context aware. The innovation life cycle becomes more efficient due to user centered and co-creation. An LIL can be sustainable and competitive for a longer time, based on contextual awareness.

## 2.3 How does Living Innovation Laboratory work?

In LIL, the innovation is generated from five phases.

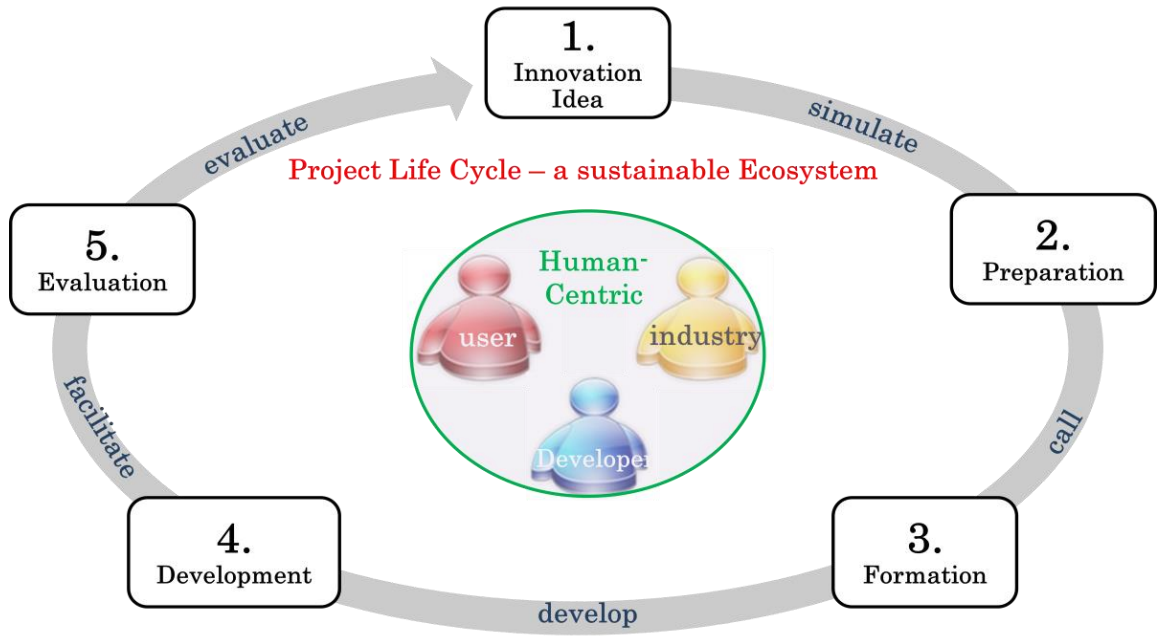

Figure 2.2: Five phases of innovation life cycle in Living Innovation Laboratory Model

**Phase 1** – *LIL Initialization*: Developers propose an innovation idea based on user needs and feedbacks. Ideas can come from any one, not just developers. Developers should listen to users' voice and pay attention to users' concern first. After all, the user-centered approach leads to requirements-driven innovation. That is why most successful corporations put a lot of effort on recognizing customer requirements on product and service.

Take the massive generation of Web 2.0 websites as an example. Some of them, such as blogs and twitter, give the freedom of designing and customizing the product (e.g. their blogs) to the users. Actually, the innovative idea of Web 2.0 just arises from user's demand and imagination. People always tend to broadcast themselves to enhance their reputation or share something with each other to broaden their social networks. They want a public platform in virtual world overcoming regional barriers to realize the dream of publishing their own-designed products (e.g. blogs, albums, homepages, and so on). Thus, Web 2.0 emerged to satisfy people's expectation, and meanwhile, brought a fantastic business model to IT industry and introduced us to a new era of world-wide Internet.

There are two aspects of this phase: 1) capture the ideas and input from a larger population, 2) understand and evaluate technology use in a specific situation. LIL enables the interaction with users, which distinguishes the LIL approach from other more traditional supplier-customer partnerships or cross-disciplinary approaches seen previously.

**Phase 2 –** *LIL Preparation*: As outside parties (e.g. companies or research organizations) who are interested in the innovation proposal invest funds on it as the role of investor, partnership or sponsorship is establised. Conditions for future research development in later stages are being set in this preparation phase.

The investors may first explore the information (i.e. revelant documents, presentations or proposals) about all projects which are being or going to be developed in the LIL , compare and choose the project which is most valuable to them. Once the target is found, they may approach the respective developing team and seek the opportunity of collaboration in the term of sponsorship or partnership. This is the first and very important step for outside parties to join the LIL project. From now on, they can oversee the project, review and give comments to faciliate the development process.

To get familiar with the new market, some survey and study is a must. It may require a certain amount of time and efforts, but it's worth it. The preparation may help LIL to define the problem, find the corresponding solutions and strategies against the potential outbound challenges, such as struggling in new market, contextual diversity, risk-assessment, and problems with scalability or integration in the future.

**Phase 3 –** *LIL Formation*: In this phase, the project is expected to attract enough investment and talents with different ideas and comments. As a project team is formed and a completed set of product requirements is finalized by the team, the project is ready to start.

Usually, project managers need to spend a lot of time and effort to search and recruit the right persons to join the developing team. Nowadays, there are more and more young researchers who are actively pursue their careers, thereby an open project may easily attract a group of talented individuals. It is also a great opportunity for them to practise their research skills and enhance their own portfolios.

Meanwhile, the detailed product design should be specified, based on the previous research. The design should be approved by development team and investors or clients, before it proceeds to the development stage.

**Phase 4 –** *LIL Development*: This phase is closely connected with the provision of user centered innovation products or services that realize the initially proposed idea, until the end products or services are released to market for testing and using. The project development process is a **co-creation** process.

Co-creating new applications and services is realized by LIL by providing bilateral access, through which users may reach the new and emerging services and meanwhile the developing enterprises may receive feedback. In the past two decades, the requirement-driven design of products and services has been universally acknowledged. Some developing enterprises in certain industries have tried to integrate users and stakeholders into product and service development, and surprisingly gained great success. For example, in Procter & Gamble, its product development processes are opened to key stakeholders, in hope of improving the acceptance of their products. Afterwards, their innovation success rate has grown by 200% in just two years, while the R&D expenditure has reduced by 3.4%. Such a case that company allows end-users and stakeholders involved in the development of a new product or a service is just according to the LIL concept. Co-creation does not only mean the collaborative but also multi-contextual environment, just like the real world.

There are two aspects of the co-creation concept in LIL:

One is related to **user involvement** in co-creation. They are engaged in the innovation process either as development sources or innovation sources, through generating contents and extending toolkits or even in an entrepreneurship role generating radical innovations. The other one is related to **industry involvement**, including academia, which is an evolution of technology transform units in universities, and city innovation promotion agencies. They may easily monitor the development process by reaching the update information about development progress and user feedbacks which is usually keeping updated online. They are also welcomed to provide feedbacks so that the project can keep going on towards the right direction parallel with their expectation. In this way, the LIL offers a channel for industry (large firms or SMEs) to actively interact with developers who are technology providers, product suppliers and project developers. For example, most of LIL in the formula of Goal Net Design, although not all, are closely coordinating with industry via Public-Private-Partnership. They are therefore usually relatively small organizations that play coordination roles between academia, companies, public agencies and other organizations, while providing a series of services either directly or through close partnership with other companies.

In the end of the development process is product deployment. Product testing should take place just prior to it. It is a common and necessary step to ensure usability and acceptance.

**Phase 5 –** *LIL Evaluation*: In this phase, developers will review the project, for example, in the form of research paper based on users' feedback collected. The analysis will help to improve the next round of research and dvelopement.

First, the team collects, records and organizes all information about project results, including short-term outputs (immediate results of activities, or project deliverables), and immediate and longer-term project outcomes (changes in behaviour, practice or policy resulting from the project). The data can be analyzed and used to answer some key questions like:

- What progress has been made?
- Are the expected outcomes achieved? Why?
- Is there any way that project activities can be refined to achieve better outcomes?
- Do the project results justify the project inputs?

It is a systematic way to collect, analyze, and use information to answer questions about projects, policies and programs, particularly about their effectiveness and efficiency. In both the public and private sectors, both of stakeholders and developers may want to know if the projects they are funding, implementing, voting for, receiving or objecting to are actually having the intended effect, and answering this question is the job of an evaluator. An additional suggestion is that the project planning stage is the best time to identify desired outcomes and how they will be measured. This will guide future planning, as well as ensure that the data required to measure success is available when the time comes to evaluate the project.

Second is to capture the user feedback. Users' voices should not be ignored but should lead the designing and development process of the product. An appropriate rewarding and incentive system may be needed to secure pay-back to all actors involved and encourge active user participation. The feedback helps slove frustrations among users and improve the product in general. The involvement of users in the project gives each user a feeling of participation. Moreover, user feedback may give birth to new ideas as it can give plenty of ideas on how to improve existing projects and maybe even fresh ideas for the following round of development and research. Refining the project based on user feedback is an iterative process. The developers would collect and anaylze the feedbacks in research papers or product documentations, whcih facilitates the futher improvement and the next round of development. Again, the key aspect in LIL to differentiate it from other cooperation's, clusters etc is the user involvement.

In this way, the development team may be able to capture users' insights, prototype and validate solustions in real life contexts. A complete and impersional review on the existing product base on user feedback is expected to give an overall evaluation on the project and then inspire the futher improvement or the next round of development.

## 2.4 Roles in Living Innovation Laboratory

There are mainly three groups of participants involved in the innovation life cycle in LIL. They have different functions and characteristics and altogether co-create the project in an open manner.

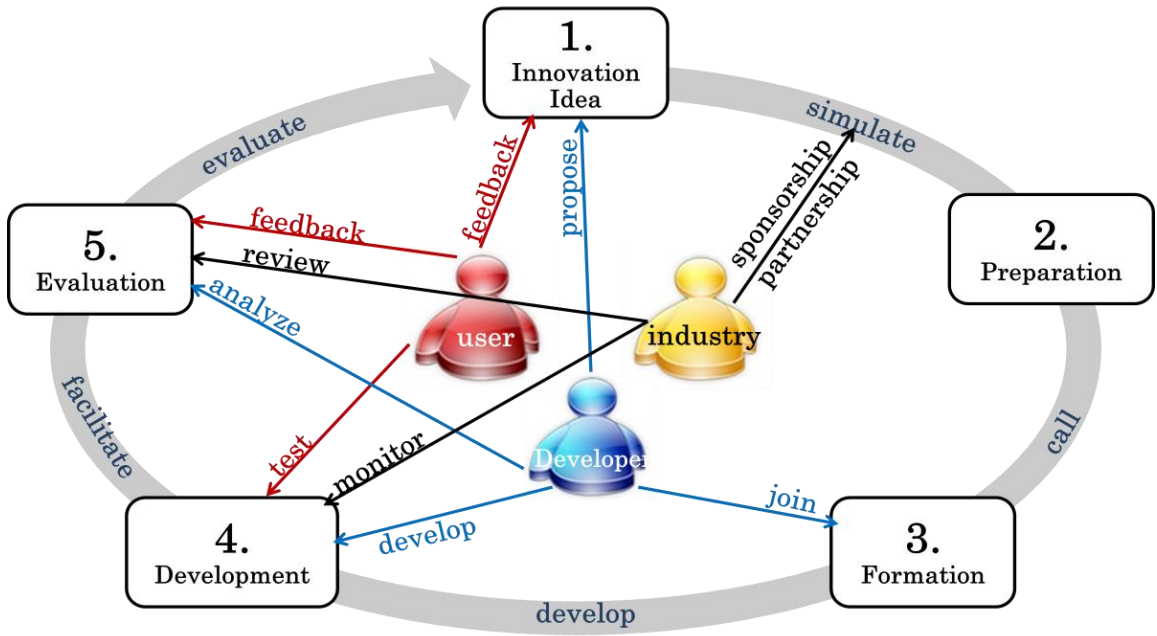

Figure 2.3: Roles in LIL and their functions

**User**

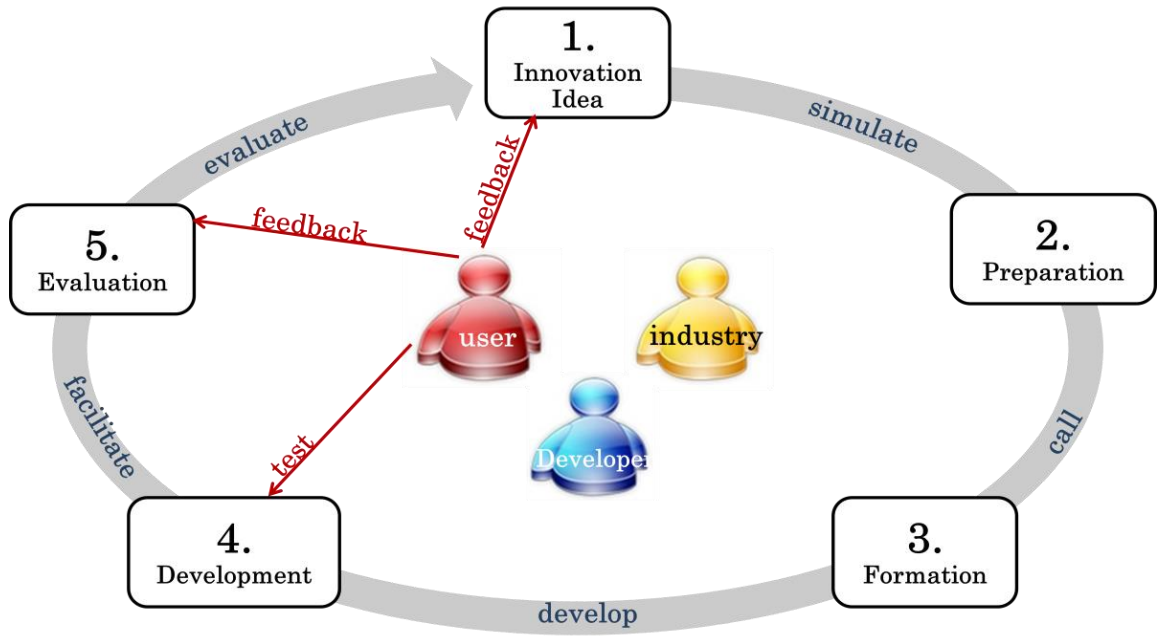

Figure 2.4: Users in LIL and their function

A successful innovation should bring some value to users. A user centered environment involves actively and interactively users and their needs in those collaborative innovation processes, far beyond only focus groups or mass surveys.

In LIL, users are organized as a pool of voluntaries while others recruit them on a project basis. By involving in the LIL innovation system, a big group of private persons (citizens and/or consumers) become a source of ideas and innovations, whose feedbacks can anytime influence the product development. They may propose, configure and invent the exact product to fit their needs. It does not only give the user a much larger freedom to innovate, but also provides a valuable feedback to the supplier. Instead of capturing needs, the development focus is changed to conveying the limitations and characteristics of the product concept towards the user as well as producing the product.

As the figure 2.4 shows, users participate in three phases described as following:

- In *Phase 1: LIL Initialization*, users show their requirement and expectation which may inspire some innovation ideas. The developer may capture users' needs through user observation and find what users want through surveys or experiments. This user-centered approach leads to requirements-driven innovation. Another way is that users may directly contribute some innovation ideas based on their own needs and wants.

- In *Phase 4: LIL Development*, users co-develop the product with developers. For example, they may help developers to conduct use case studies, raise their requirements which may change with time, or participate in user testing in the real life context before the product's delivery.

- In *Phase 5: LIL Evaluation*, users are free to provide feedbacks, comments and suggestions. Those data can be collected from users in various ways, including survey, experiments and observation. The cross-analysis of user feedback allows developers to evaluate the current project and boosts the next round of development and research. It may even inspire some new ideas to facilitate innovation expansion.

## Developers

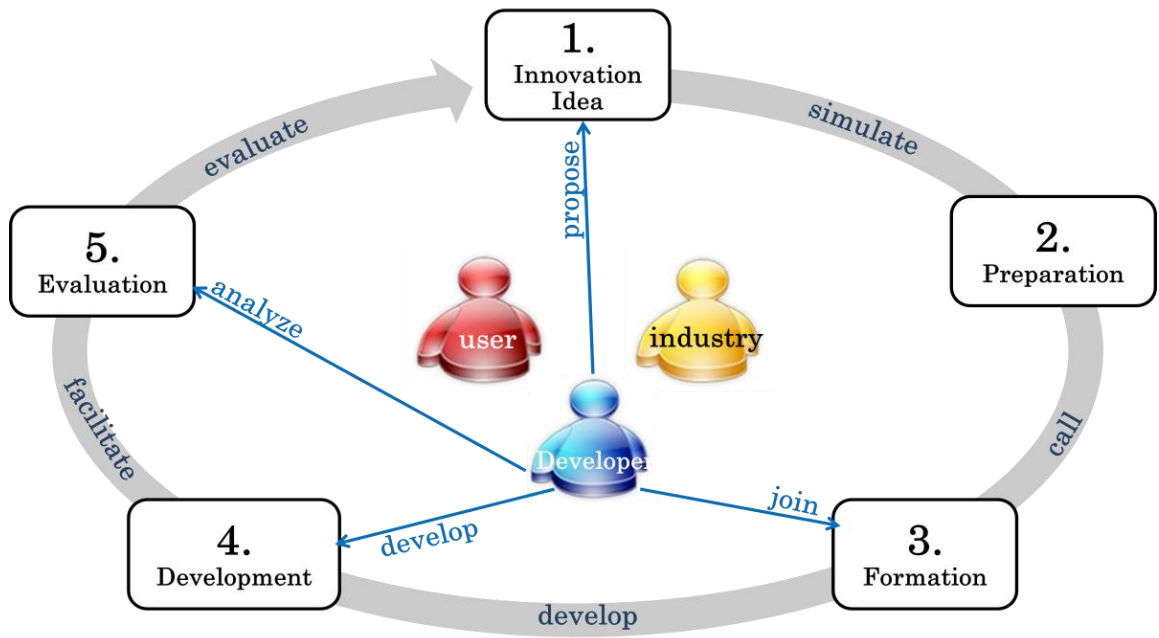

Figure 2.5: Developers in LIL and their functions

As the figure 2.5 shows, developers participate in four phases described as following:

- In *Phase 1: LIL Initialization*, they propose innovation ideas according to user requirements and society needs. The proposal is open for deeper discussion and further improvement in the future. Anyway, it initializes the project in the first place and attracts the research funding in the next stage.

- In *Phase 3: LIL Formation*, developers from different fields and with different skills come together to form a developing or research team or join an existing team. They are all interested in and have passion on developing the project. They, of course, are able to contribute in the development in any way so that they may be qualified to become team members and recruited by the project investigators.

- In *Phase 4: LIL Development*, in order to better develop the product in a co-creation concept, they coordinate with other parties, including not only employees and internal stakeholders, but also customers, suppliers, and related external stakeholders and communities. Co-creation does not only mean a trend of jointly creating products, but also describes a movement away from customers buying products and services as transactions, to those purchases being made as part of an experience. LIL provides a platform for developers, active users and shareholders to share, combine, and renew each other's resources and capabilities to create value through new forms of interaction, service and learning mechanisms. This "full theory of interactions" goes beyond the existing forms of co-creation of the customer experience and co-creation of products and services. Developers should take note of users' feedbacks and needs to adjust their production plan as well as shareholders' requirements and suggestions to satisfy their expectation, any time during the development process. The personalized products are allowed to be specially designed to suit users' needs. Prior to product or service delivery, user testing is compulsory, so that the last round of refinement can be done before the official release of the product or service.

- In *Phase 5: LIL Evaluation*, developers collect data of project results as well as feedbacks from user and industry. Upon analysis, evaluation may be done in the form of published paper, internal report and so on. As project evaluation is the control of the planning and implementation of project activities with regard to the objectives to be achieved, the assessment and documentation process should take place at two levels at least. At the first level, it is necessary for the project team to collect all project results and assess them against the overall objectives of the grant-making project. At the second level, the feedbacks from users and shareholders are important here for the purpose of determining the success of the project and better refining the project in the future.

Industry

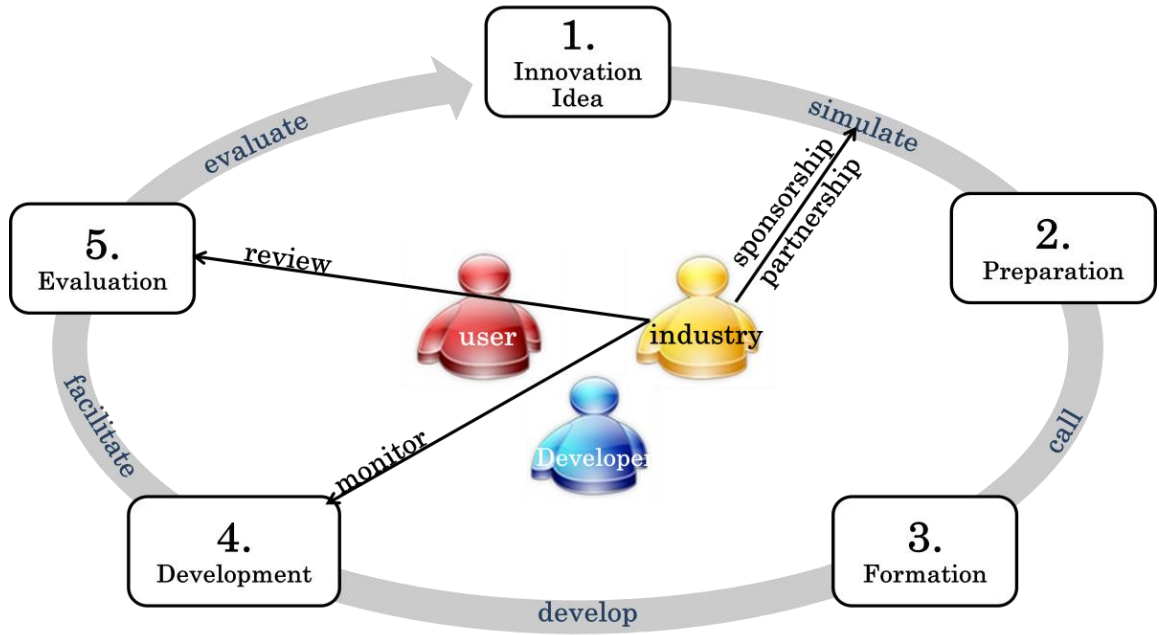

Figure 2.6: Industry party in LIL and their function

As the figure 2.6 shows, industry organizations participate in three phases described as following:

- In *Phase 2: LIL Preparation*, industry organizations, who are seeking opportunity to invest or participate in any project in the related fields, plays the key role. Project initiators should publish the project proposal and open it for investigation and discussion, which may attract research funds from industry. For example, they may display the project information online or propose in research conference. In a word, industry organization should easily reach the project information and approach the project investigators. There are lists of projects open for investment. Industry organizations are free to choose the one they are interested in, and then join the respective innovation life cycle as sponsors or partners. For example, provide research funds to establish sponsorship or other resources, such as groups of users, to establish partnership,

- In *Phase 4: LIL Development*, LIL defines that the "co-creation" development process does not only involve end-users but also shareholders. Industry organizations are able to monitor the project against their business objectives. The important views from their perspectives are to be considered as a very helpful resource of development. After all, the most important business objective of the project is to meet industry requirements; otherwise, the project is impossible to survive.

- In *Phase 5: LIL Evaluation*, industry organizations are to examine the project outcomes and user feedback. On one hand, as a project nears completion, it is a good opportunity for the organizations involved in the project to take stock of what has been done and to document this innovation production. On the other hand, those sponsors and partners may become a source of evaluation, as their feedbacks are also to be collected and analyzed for the future improvement of the product or service.

In conclusion, as the research approach in LIL is human-centric, there are always different groups of participants involved in different phases of innovation life cycle.

## 2.5 Summary and Discussion

LIL is a research concept which introduces a collaborative working environment for users, developers and industry party to co-develop user centered projects in an open manner.

The three main characteristics of LIL are user centered, co-creation, and context aware. First, LIL involves end users into development of new applications and services by providing bilateral access, on the one hand, of the consumer to the new and emerging services, and on the other of the developing enterprises to their feedback and contribution [3]. Second, LIL encourage the co-creation of different parties including users, developers and industry party. Third, the multi-contextual real-world environment is highlighted in LIL, which may help an innovation survive in the changing world.

In a word, LIL presents a user-driven research infrastructure in adoption of a systematic co-creation approach integrating remote resources, various expertise and innovative ideas together. It provides a sustainable ecosystem for a multidisciplinary team involving users and developers to work together in an open-ended experiential environment.

# Chapter 3:

# The New Living Innovation Laboratory

After reviewing the concept of LIL in the last chapter, this chapter described the bottleneck with the existing LIL. Some possible solutions are proposed, including unobtrusive user involvement, massive co-creation, and predictable context aware. It results in a new generation of LIL with new characteristics and more advantages.

## 3.1 The refinement of Living Innovation Laboratory Model

Albert Einstein said, "if I had 20 days to solve a problem, I would spend 19 days to define it." So, before we refine LIL, let us clarify what an innovation means. Basically, there are two levels of innovation in terms of breakthrough degree:

1. Incremental Innovation: It happens when a company needs to keep their products competitive and maintain a certain amount of market share over time. They seeks to improve their products by making them better, faster, and cheaper with lower cost. Incremental Innovation will not expand the existing market boundary or uncover any new market.

2. Disruptive Innovation: It was first introduced by Clayton Christensen introduced in his classic book *The Innovator's Dilemma*. When an innovation brings a different set of values beyong users' expectation but indeed better people's life, and eventually create a new market or even takeover the existing market, we call it as "Disruptive Innovation". It can be a new techology, product or service, which unexpectedly makes a revolution in the industry.

The difference between these two levels of innovation is that disruptive innovation is unexpected and create a new market while incrementary innovation caters to clearly defined user requirement within the existing market. For example, iPhone 6 is an incrementray innovation in smart phone market. The innvetion of the first iPhone is a disruptive innovation which people never imagined before. It created a new market called "Smart Phone" and untimately tookover the previous market called "Feature Phone".

After understanding the two innovation levels, let us see whether LIL 1.0, which we talked about in literature review, can achieve disruptive innovation. If not, how to break through limitations in LIL 2.0?

### 3.1.1 Unobtrusive User Involvement

**Why Unobtrusive User Involvement?**

Traditionally, to recognize user requirements, the development team would design questionaries and invite users to participate in the way of focus group, interview, survey and so on. There are two drawbacks:

First, the pre-design questions may direct users towards the area which the development team expect. For example, a survey about how to improve Nokia 3310 (feature phone) may contain the questions mostly about press buttons and liquid crystal display. It may limit the divergent thinking of users.

Second, in tens years ago, people had no concept of other advanced types of hand phone besides feature phone. In this limited context, it is impossible for the idea of smart phone to come from user requirement.

Hence, while the development team empathized "user centered" in LIL and link their products too closely to the users, it may result in creating incremental innovation [1]. Incremental innovation can improve the existing products into the new version. But it is hard to discover the new market just based on the traditional "user centered" research method. So, is there any other way to discover what users actually need but cannot think out themselves? A new concept called "Unobtrusive User Involvement" was proposed, which may help us to head to disruptive innovation.

Steven Jobs described that, the real innovation is to give people what they never dream about but must feel so exited when they are given, like iPhone. Before iPhone, we interacted with hand phone via keyboards and PDA via stylus pens. No one realized that we had better use fingers. How did Steven Jobs find that? Because human desires are sometimes revealed inadvertently in daily life, instead of a user study environment where people are well aware of their roles. Through the big data collected from not only designed user stuy but also trivial logs, Jobs found that when people interact with machine via fingers, they would feel more comfortable and confident due to the immediate response. This discovery from unintended data, instead of user study, is very important to Jobs. He insisted to develop and release the first touch-screen smart phone in the world. People were passionate about this unexpected device. Afterwards, all kinds of smart phone were invented with touch screen. A new market was created and finally dominated the hand phone industry. That is just a typical example to show why LIL needs Unobtrusive User Involvement (i.e. UUI).

**What is Unobtrusive User Involvement ?**

UUI provides a way to make use of collected user data no matter users are aware or not as long as legally. The traditional user study in LIL, such as focus group and survey, collects user feedback when they are aware of it, while the user study by using big data summarizes user behavior when they are unaware of it. All of user data are useful to create disruptive innovation in user centered LIL.

**How to achieve UnobtrusiveUser Involvement?**

Nowadays, user behaviors generate a hug amount of data in the cyber world, so called "Big Data". By analyzing the **big data**, we may be able to discover the hidden demands and nature feedback from users.

Take Joycity, a famous shopping mall in Beijing, China, as an example. There are Wi-Fi points all over the mall, so that customers can connect to Internet via it anytime. The big data of Wi-Fi connection generated from user behaviors is collected and analyzed by Joycity. Based on it, Joycity draw a hotspot map to illustrate which path most customers love to go along with, that is, the pefect path for shopping. If Joycity interviewed with customers and asked them directly, customers might not record or realize their favourite path which could be only reflected in their daily behaviors. Based on this hotspot map, Joycity re-designed the store layout in the shopping mall, in order to provide a better shopping experience to customers. Meanwhile, Joycity also tracked the sales performance of the stores before and after the re-design, in order to prove whether their analysis and design were correct or not.User feedback, represented by the sales performance of the stores, would show whether the re-design met users' expectations, even though users never explained their expectations directly. In this way, Joycity used UUI and design *by* user approach to conduct a LIL project successfully. Thanks to big data, users can effectively and efficiently co-develop and evaluate a LIL product, though they do not notice it.

**Advantages about Unobtrusive User Involvement?**

On one hand, UUI can discover user requirements <u>effectively</u>, no matter users notice their requirements or not.

On the other hand, UUI can save cost for LIL to achieve "user centered". Instead of spending a lot of money and time to reach enough users to conduct survey, now LIL can <u>easily and quickly</u> gather tons of users and their feedbacks through UUI.

It is worth to mention that big data techology, such as data mining and big data analysis, is quite helpful for LIL to carry out UUI.

### 3.1.2 Massive Co-creation

**Why Massive Co-creation?**

In LIL 1.0, co-creation refers to the collaboration of three parties (user, developer, and industry party) in order to jointly produce a mutually valued outcome [8]. That is, the development team, consisted of some team members, invite a few shareholders, sponsors and clients (industry party) as well as a certain number of users to participate in the project. The scope of co-creation is limited, as the number of co-creators is not big enough. The user feedback may only represent the position of a small group of people. It is better to involve more people, in order to avoid bias and established thinking pattern. Thus, a new concept called "Massiver Co-creation" emerges in LIL 2.0.

**What is Massive Co-creation ?**

Massive Co-creation (i.e. MCC) refers to the utilization of the crowd (manpower and intelligence) to solve a problem together or co-develop a product. For example, Apple Group developed and released one generation of iPhone per year, on average. In contrast, Google set Android as an open source and call for the world-wide crowd to improve it together. It turns out that MCC is more productive than one company (Apple Group), as hundreds of generation of Android phone are released every year. Despite the uneven quality of products generated from MCC, we can still see some surprising products with high quality, such as Samsung and MIUI.

In a word, LIL 2.0 tends to invlove much more people in the co-creation process, compared with LIL 1.0. There are three approaches to achieve it: crowd sourcing, crowd funding, and crowd testing.

**How to achieve Massive Co-creation?**

**Crowd sourcing** means the company invites the crowd to to come with ideas or solutions, in the hope of finding the best ideas or solutions which will be rewarded [9]. There are two types of crowd sourcing (Brabham, 2010):

    i.    abstract creative ideas from the crowd (crowd sourcing creative tasks) which can be applied to the initialization stage of LIL.

    ii.    gather the crowd to solve a problem (crowd sourcing complex tasks) which can be applied to the development and evluation stage of LIL.

Through crowd sourcing, LIL may attract a large number of diverse people with their ideas and talents to co-create one project online or offline, despite the region limit.

**Crowd funding** is the practice of soliciting financial support for a project by raising monetary contributions from the crowd, typically via the online platform [10]. As the crowd become shareholders of LIL, they have more movitation to participate in the co-deveopment sections without confidental issue. In 2013, the crowd funding industry grew to over $5.1 billion worldwide [11]. Since more and more people accept this way to join in a project, it is a good time to adopt it in LIL. It does not only bring more funding for LIL formation, but also faciliate the following massive co-creation in LIL development and evaluation. Furthermore, the crowd as shareholders must be more than willing to become the customers of the product they jointly invested and developed. The large number of inherant customers will secure the success of the product in terms of sales performance and thus reduce business risk. For example, Yule Bao under Alibaba Group is a crowd funding platform for entertainment products, such as movies, TV series and games. Millions of users can become movie investors by investing 100 RMB or

above and then have the rights of casting, selecting directors and so on to co-produce a movie. In the end, they may get discounted tickets to see the movie in cinema and around 6.5% annual return from the box office earnings. Yule Bao attracted 240,000 user participants with 2 days and raised 7.3 million RMB for 5 movies and 1 online game. As we can see, through crowd funding, LIL can easily and quickly involve a large number of end users as shareholders as well as active co-creators of a project.

**Crowd testing**, as a form of crowd sourcing, gathers the professional and unprofessional testers with different backgrounds and from different places to test a product under diverse realistic platforms to ensure the product is reliabe, bug-free and able to meet user requirements [12]. In LIL 1.0, a targeted group of users and shareholders are invited into testing session. It differs from crowd testing in which the testing is carried out by an unlimited number of diverse users and shareholders (if crowd funding). More bugs could be found and fiexed within a shorter time; and more user feedbacks could be collected for further improvement.

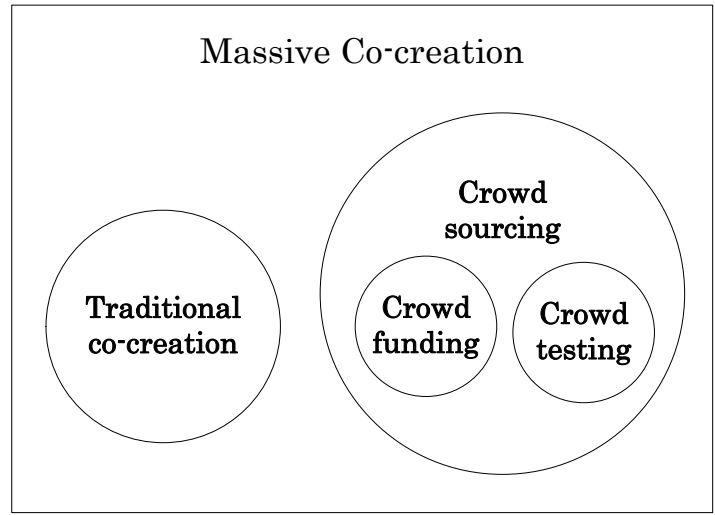

Figure 3.1: The approaches to achieve Massive Co-creation

By integrating crowd sourcing, crowd funding, and crowd testing into LIL, co-creation can be further enhanced to massive co-creation.

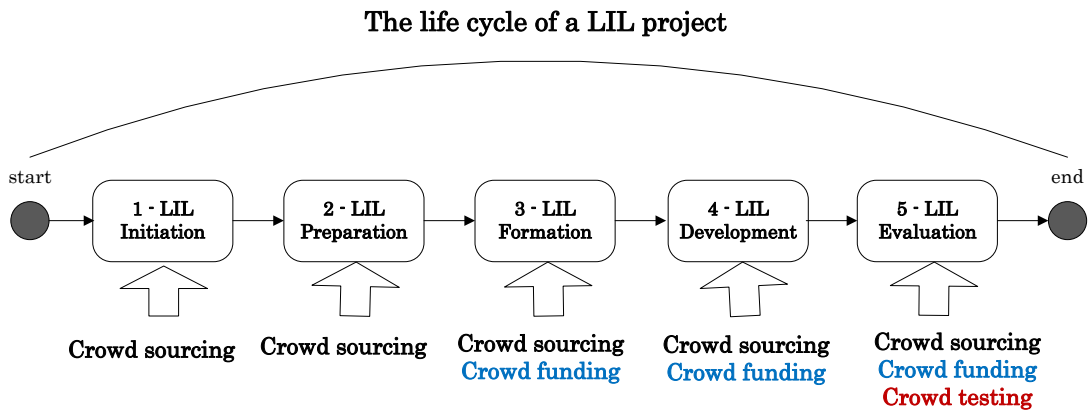

Figure 3.2: How Massive Co-creation affects the life cycle of a LIL project

The figure above shows that how MCC infuses into different stages of LIL. Crowd sourcing gathers a wider range of collective intelligence together for the initialization, preparation and development of a project, throughout Phase 1 to 6. Crowd funding brings in investment in Phase 3 and shareholder participants in Phase 4 and 5. Crowd testing enlarges the testing group and evaluation scope in Phase 5.

**Advantages about Massive Co-creation?**

The major difference between MCC and traditional co-creation is that MCC makes use of mass talents instead of a target group of people. MCC broadcasts problems to the public and calls for contributions to solving the problem (Howe, Jeff, 2006). It definitely improves the effectiveness and efficiency of the project. Hence, LIL 2.0 with MCC can develop and validate a project quickly and cheaply based on larger sample sizes.

### 3.1.3 Predictable Context Aware

**Why Predictable Context Aware?**

The word "innovation" is derived from Latin word "Innovare" which means "in(within) + novare(change) ". Innovation does not only mean "some changes", but also "some changes happening in and due to the changing context".

In 2010, Tecent, as one of the largest Internet companies in China, noticed the rising market of mobile social network. It decided to develop an APP specially for mobile social network, even though it had a smiliar APP called "Mobile QQ" already. After three months, Tecent released "WeChat", which gradually grabed the users from Mobile QQ and MiTalk, and eventually dominate the market of mobile social network. Now in China, more people tend to send messages via WeChat rather than SMS, which disrupted the previous telecom market (i.e. disruptive innovation). Later, the three giant companies in Chinese telecom market (i.e. China Mobile, China Telecom, and China Unicom) tried to get back their users by releasing similar APPs but all failed.

In this case, the most changing context is that people tend to use mobile chatting APPs instead of SMS. Most of companies did not predict this change, like the three giant companies in Chinese telecom market. They only noticed the change after seeing it. But it is too late, as the market has been occupied already.

LIL 1.0 requires us to be aware of and adapt to the changing context, like what the tree giant companies did. But it seems not enough for companies to survive in the rapidly changing world now. That is why a new characteristics called "Predictable Context Aware (PCA)" is proposed in LIL 2.0.

**What is Predictable Context Aware?**

The concept of PCA requires LIL to predict the upcoming change and prepare the solution and strategy before the context actually changes. The changing context includes expanding or shrinking market, emerging technology, competitors, user demands, and so on. In the case above, Tecent foresaw the change, acted very fast to catch up the change, and eventually became the winner.

**How to achieve Predictable Context Aware?**

It is actually harder to estimate the changing context than just monitor it. Again, **big data** needs to be used. Big data is a set of techniques and technologies to uncover large hidden values from large datasets that are diverse, complex, and of a massive scale [14]. Usually, the trend of the changing is just hidden in the integration of counterless trivial data.

In the middle of 2013, a groupbuy website "Meituan" seeked their strategy to survive in Chinese groupbuy market. They collected the big data about groupbuy users' behaviors in PC and mobile. At that time, over 70% groupbuy sales came from PC. But after the big data analysis, they found that in the near future most of groupbuy users would move from PC to mobile. Because, once a user purchased via mobile, it was more possible that he or she would remain in mobile and make the future purchase via mobile instead of PC. Plus, the simply purchase steps in mobile APP shortened the decision making process of the buyer, so that mobile shoppers tended to buy more via mobile based on quick decision. Hence, Meituan predicted that the ratio of groupbuy market shares in PC and mobile would be totoally inversed by the end of 2013. Their strategy was to allocate more resources on mobile APP than PC website. The fact is that in the end of 2013, around 70% groupbuy sales generated in mobile, no longer PC. As Meituan changed their strategy on time, they got survived in 2013. In this year, the number of groupbuy companies decreased from 1478 to 213. Now, Meituan becomes the largest groupbuy company in China with 55% market shares (EnfoDesk, 12/11/2014).

This is a typical example of using big data to anticipate the upcoming changes through big data. The detailed big data metholodogy will not be illustrated in this papter. But we should keep in mind of the big data approach to predict the context.

**Advantages about Predictable Context Aware**

Long time ago, we make the judgement based on our experience; now, we use big data to verify whether our judgement is reasonable or not. Through big data, we can make the more accurate anticipation than before. If the contextual change can be predicted, we can adjust the project to adapt to the change before it happens and ahead of potential rivals.

## 3.2 The advantages of the new Living Innovation Laboratory

LIL 2.0 is distinguished from LIL 1.0 in terms of the three characteristics, summarized in the table below. Based on them, LIL 2.0 gains significant advantages over LIL 1.0.

| Characteristics of LIL 1.0 | Characteristics of LIL 2.0 | New methods that LIL 2.0 adopts | Advantages of LIL 2.0 over LIL 1.0 |
| --- | --- | --- | --- |
| User centered | Unobtrusive User Involvement | Big data | Effective, Efficient, Lower risk |
| Co-creation | Massive Co-creation | Crowd sourcing, Crowd funding, Crowd testing | Effective, Efficient, Lower risk |
| Context aware | Predicatable Context Aware | Big data | Active, Lower risk |

Table 3.1: The differences between LIL 1.0 and LIL 2.0

UUI, MCC and PCA gather massive intelligence from the world-wide crowd to co-produce a product or service in a predictable context. With thee three characteristics, LIL 2.0 becomes more effective and efficient with lower business risk, compared with LIL 1.0. In addition, LIL 2.0 continues to have a deep insight of the market through big data, actively uncover the future trends, and eventually make itself sustainable in the changing world.

## 3.3 Summary and Discussion

LIL 1.0 has three characteristics, that it, user centered, co-creation, and context aware, which may not be perfect enough. First, the inspiration in LIL 1.0 usually comes from what users require, which may result in incremental innovation. Because users may not know what they actually want, not explain very well, or be misled by the pre-designed questionnaires. Second, the participants in co-creation are pre-selected, but we actually need more diverse sample. Third, it may be too late if LIL only act after the contextual change happens. So, LIL 1.0 should be improved with new characteristics: UUI, MCC, and PCA.

UUI tries to get more users invloved in LIL through the traditional user study and big data approach. The users may act more naturally when they do not notice the test bed environment. A large number of diverse user data is collected, which results in more creative ideas and userful suggestions.

MCC brings in more partipants to make LIL more productive. Crowd sourcing extends the co-creation team from a limited group of people to the global scale.Crowd funding helps to raise investment and facilitate LIL formation and evaluation. Crowd testing allows the world-wide testers to joinly evaluate a product within a shorten time and despite the region limit.

PCA gives LIL more confidence to face the changing world and sudden challenge. Big data can help LIL to anticipate the future trends by digging out the hidden values from huge datasets.

As a result, the concept of LIL 2.0 emerges, powered by UUI, MCC and PCA. It is designed to make the innovation process more effective and efficient with lower risk. In Chapter 5, LIL 2.0 would be applied to real cases, to show how better it is than LIL 1.0.

# Chapter 4:

# Living Innovation Laboratory Model Design based on Goal Net

So far, some methodologies have been proposed in other research papers, but all with very high-level design. No one has summarized a detailed methodology for someone who never knows about LIL to follow and then can build a qualified LIL easily. In this chapter, such a detailed methodology for LIL was proposed, based on Goal Net.

## 4.1 Introduction to Goal Net Model

### 4.1.1 The theory of Goal Net Model

Goal net is a composite goal model to facilitate a progress. Its compositions include states, transitions and environment variables [30]. A Goal Net structure is shown as follows. A goal is a desired state to reach. A transition represents a goal relationship between goals.

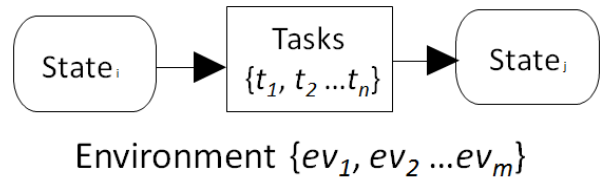

Figure 4.1: The elements of Goal Net Design

Circles and arcs or vertical bars represent states and transitions, respectively. An agent needs to go through the states for the purpose of achieving final goal. The input state is connected to the output state by transitions. To define the possible tasks an agent needs to take to achieve the goal of transiting from the input state to the output state, a task list is associated with each transition. Figure 4.2 shows a Goal Net example.

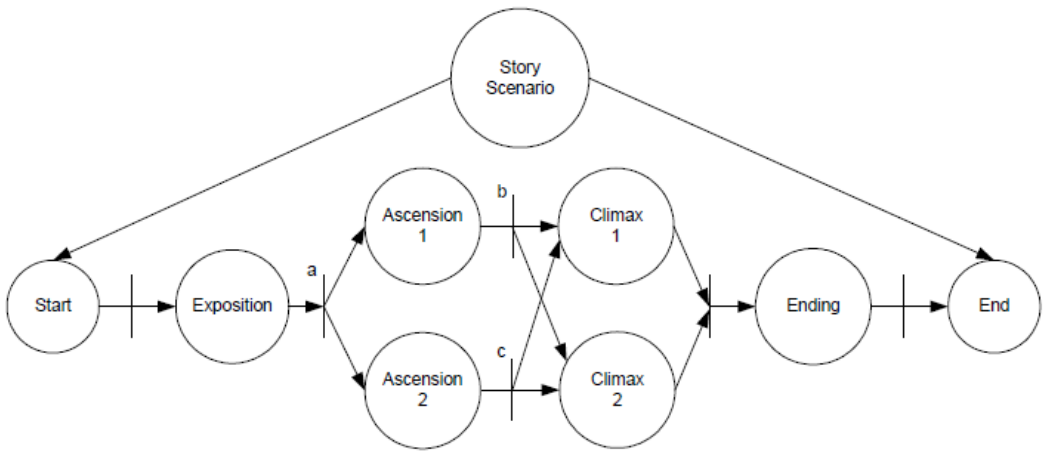

Figure 4.2: Goal Net with Alternative Storylines [40]

Two kinds of states are specified in Goal Net, which are atomic states and composite states. An atomic state is a primitive state which cannot be divided any more, while a composite state can be split into states connected via transitions. Therefore, a complex goal can be recursively decomposed into sub-goals and sub goal nets. The hierarchical structure simplifies the goal oriented modeling process with different levels of abstraction. In Goal Net, four types of temporal relations of goals are represented by transitions connecting the input states and output states: sequence, choice, concurrency and synchronization, which are shown in Figure 2.3.

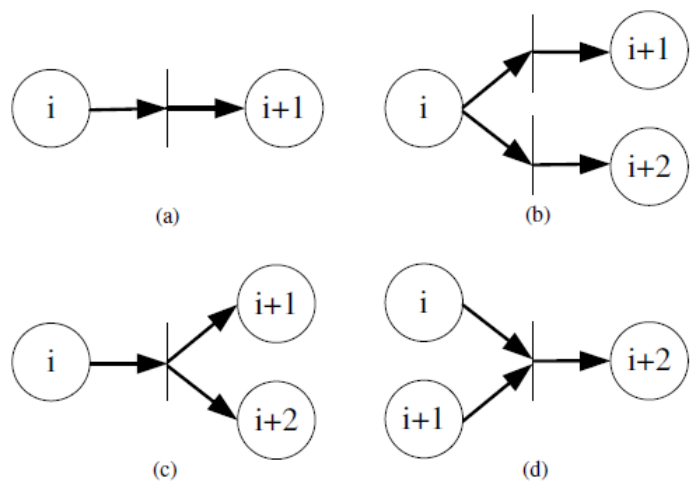

Figure 4.3: Goal Net Transitions: (a) Sequence (b) Concurrency (c) Choice (d) Synchronization [40]

The transitions have the following meanings:

- *Sequence*: A direct sequential causal relationship between input state i and output state i + 1.

- *Choice*: A selective connection from input state i to possible output states i+1 and i+2, and only one output state can be selected based on selection criteria.

- *Concurrency*: Input state i at completing the tasks, all the output states i + 1 and i + 2 can be achieved simultaneously.

- *Synchronization*: A synchronization point from different input states i and i + 1 to a single output state i + 2, and the output state can only be achieved when all its input states are synchronized.

There are two types of arc in choice situation: triangle arrows represent "or" relationship between tow triangle arrows, while diamond arrows means "and" relationship between two diamond arrows [30].

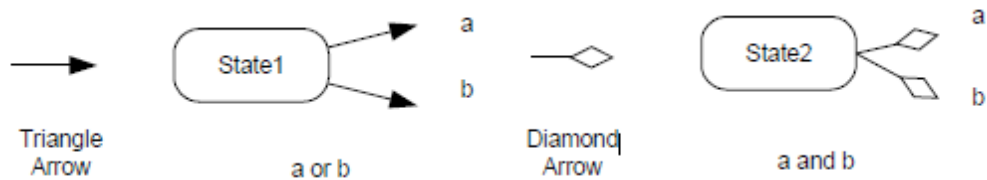

Figure 4.4: The types of arcs [30]

In a Goal Net model, a state is represented as Si, and the transition is represented as Ti.

Goal Net is a very expressive and efficient tool to model a process in an interactive context. Thus, a LIL Methodology can be developed in the form of Goal Net. For example, a composite state in Figure 4.5 represents a dynamic goal pursuing process to achieve the goal.

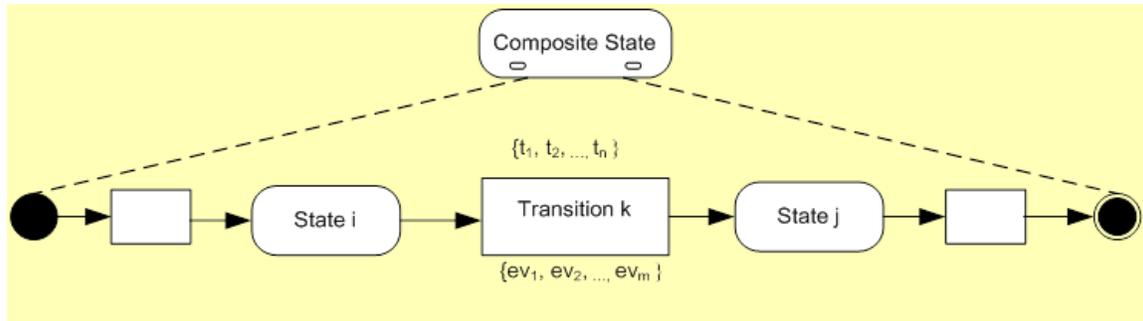

Figure 4.5: Goal Net's composite state

If the progress of a LIL project is described in Goal Net format, the project goal pursuing process can be demonstrated in Figure 4.6. A composite state consists of a process (project goal pursuing process), an initial state, a target state and a pair of branches.

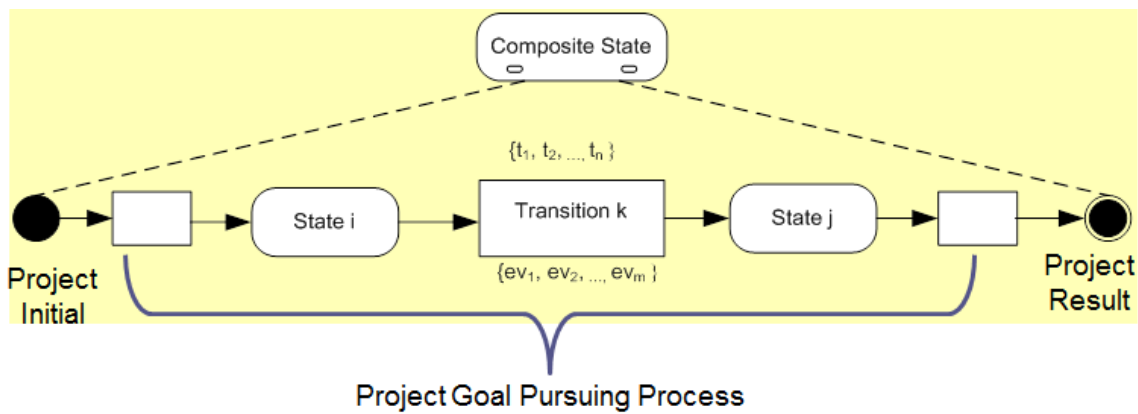

Figure 4.6: Project Goal Net process

As Goal Net is a very expressive and efficient tool to model a process in an interactive context, LIL model can be designed based on the Goal Net Model.

### 4.1.2 The reasons of using Goal Net Model

Goal Net Model is a goal-oriented modeling approach for engineering a complex and distributed system in a dynamic environment, such as multi-agent system [30]. Unlike a flowchart, it emphasizes the outsides variables which cause the system to generate different stories. Similarly, LIL is also required to be built in an open, complex and dynamic operating environment.

Besides, in a complex online game built on Goal Net Model, an agent may give different performances based on different contexts. It is like that Goal Net gives the agent an artificial brain to select different ways to pursuit the goals, where goal selection and action selection strategies are to be used. Let us assume LIL as the complex online game, the leader of LIL as the agent, and a detailed guideline to build LIL as the game storyline. Even if the leader of LIL has no experience about LIL, he or she can still build an LIL by following the detailed guideline. It is so called LIL Model Design based on Goal Net, which would be illustrated in the following section.

## 4.2 Living Innovation Laboratory Model Design based on Goal Net

The previous chapter shows that LIL model is consisted of five phases: initialization, preparation, formation, development, and evaluation.

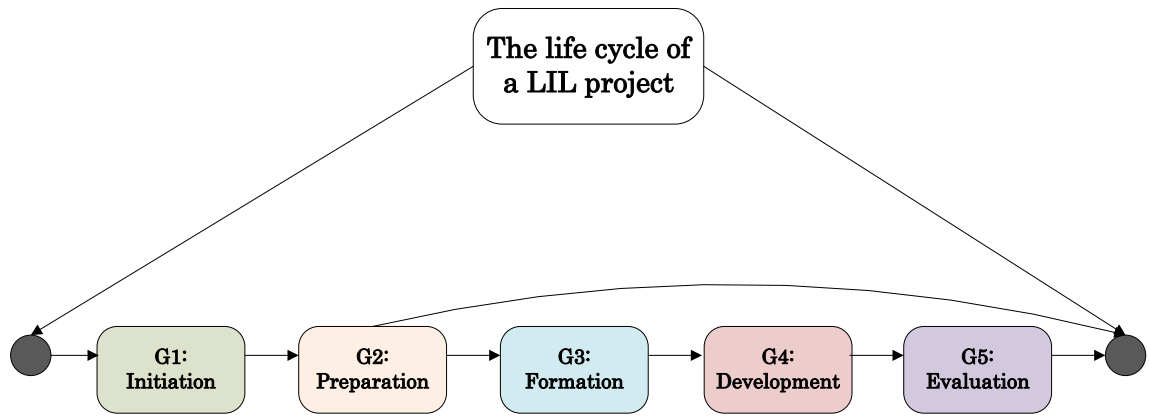

Figure 4.7: Five phases of LIL from Goal Net perspective

It is worth of mention that LIL Model Design based on Goal Net is not only a detailed and systemic methodology towards LIL 1.0 but also LIL 2.0.

**Phase 1 –** *LIL Initialization*:

(Involved parties: Industry, Developer, User. Keywords: big data, crowd sourcing)

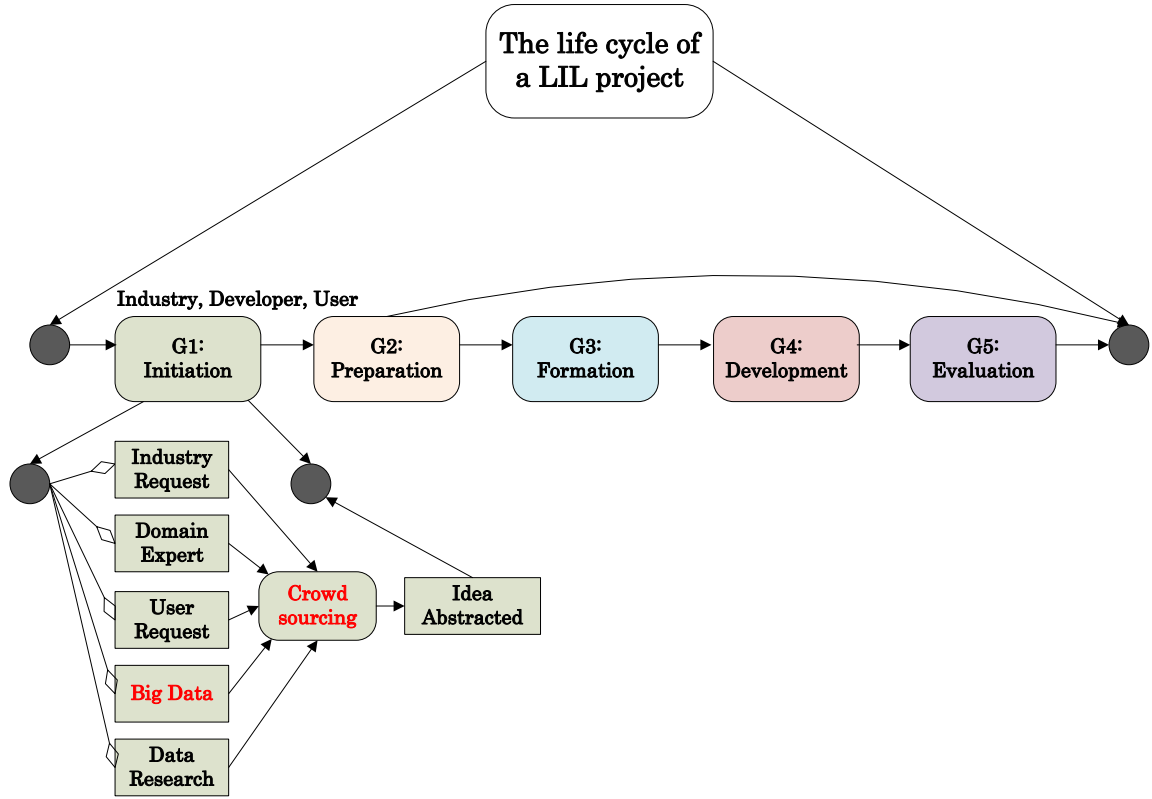

Figure 4.8: The composite goal of LIL Initialization from Goal Net perspective

The innovation-fostering requirements come from various sources, including:

- Industry Request: means the requests from shareholders, sponsors or clients, for business purpose, no matter it is innovative or not. The most common advertisement form on Internet is "Cost per Click", which is widely in search engine now. It just derived from advertiser request with no technical breakthrough. But it established an innovative Internet business model which was proved to be the most successful one.

- Domain Expert: refers to the engineers and scientists who are experts in their research field. They pay close attention to new technologies, have deep insight of innovation, and think about any possibility that a new technology and a product may combine and offer something new. They should be the most powerful source of innovation. It is just like that Thomas Edison, as a domain expert, invented electric light bulb.

- User Request: The "customer-as-the-king" model encourages users to express their requirements as they are the one who ultimately pay for the product. User involvement do brings a lot of inspiration to create or improve a product.

- Big Data: is an important approach towards UUI. By adopting big data in the initial stage, LIL would meet user requirements and even beyond user expectations. Besides, domain experts may also get inspired by big data.

- Data Research: is the research on various legacy materials in both of standard and digital libraries, in the quest of innovation [1].

Any idea from industry, domain experts, users, big data, and legacy materials can initiate an LIL project. "User request" makes the project user centered from the beginning onwards. "Big data" brings UUI and MCC to the project, so that it tends to be an LIL 2.0 project. Furthermore, the initial ideas should be refined by crowd sourcing. Because, many rough ideas are unpractical or too subjective. Through crowd sourcing, we gather the crowd intelligence to quickly verify which ideas and assumptions are reasonable and do-able. Finally, the best idea can be abstracted for LIL to further proceed.

**Phase 2 –** *LIL Preparation*:

(Involved parties: Industry, Developer, User. Keywords: big data, crowd sourcing)

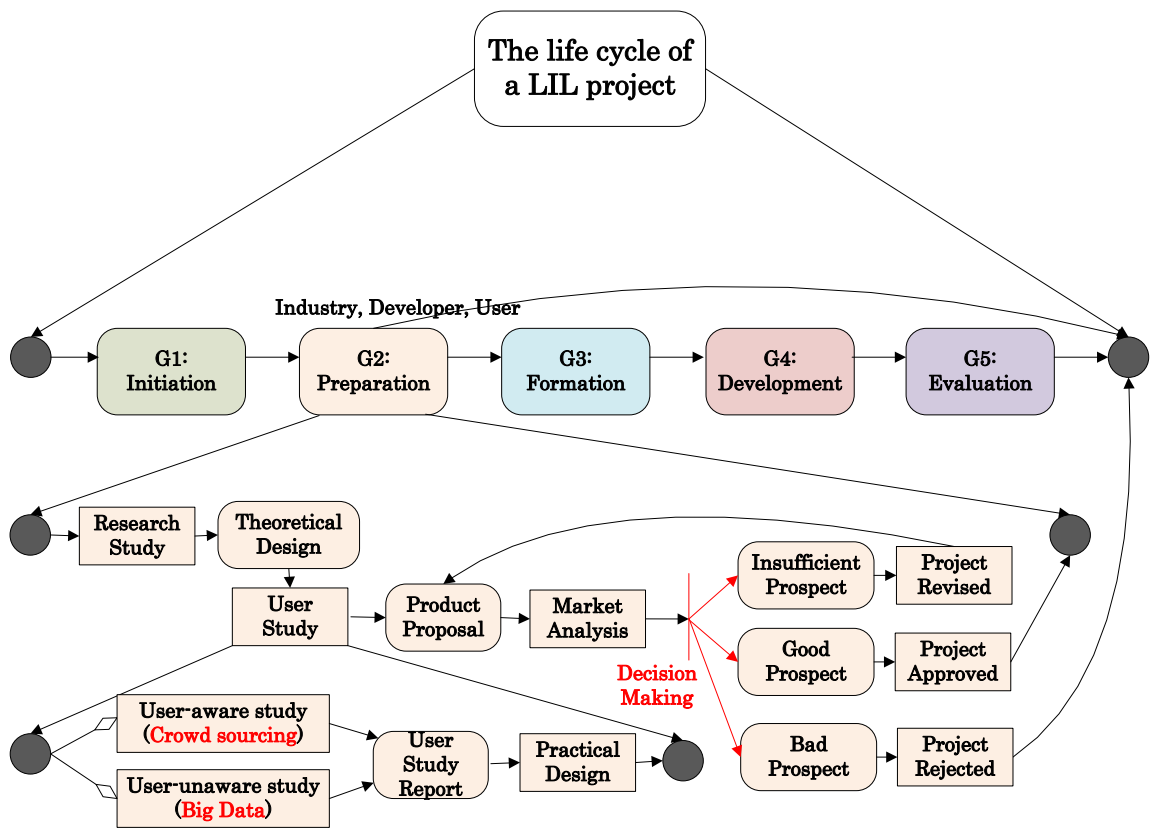

Figure 4.9: The composite goal of LIL Preparation from Goal Net perspective

It is a critical phase which can kill the project immediately if the outcome of preparation is undesirable. Research study is to find some theoretical support to prove the product is technically realizable and logically reasonable. But theoretical design may not be appealing to users, so should be followed by user study. User study is a composite goal, which is consisted of a serial of user centered research towards a practical product design. The user centered research should be conducted in both of obtrusive and unobtrusive ways where crowd sourcing and big data can be used. All user data are collected, analyzed, and composed into a user study report, which is a useful guideline for practical design as well as the most convictive evidence to show the proposed design will be able to meet user requirements. By combining theoretical design and practical design, a product proposal is ready. In order to proceed to development, we must convince the investors (industry party) that it has great market prospects. If the product has novelty but too small market, no investor would be interested and the project should be closed to re-consider other ideas. If the product has big market potential but the design is not attractive, the proposal should be revised and audited by investors again, until finally approved. In conclusion, LIL preparation is to ensure the success of the product in terms of market prospects and design quality, otherwise cease the futureless project immediately to avoid resource waste.

**Phase 3** – *LIL Formation*:

(Involved parties: Industry, Developer. Keywords: crowd funding)

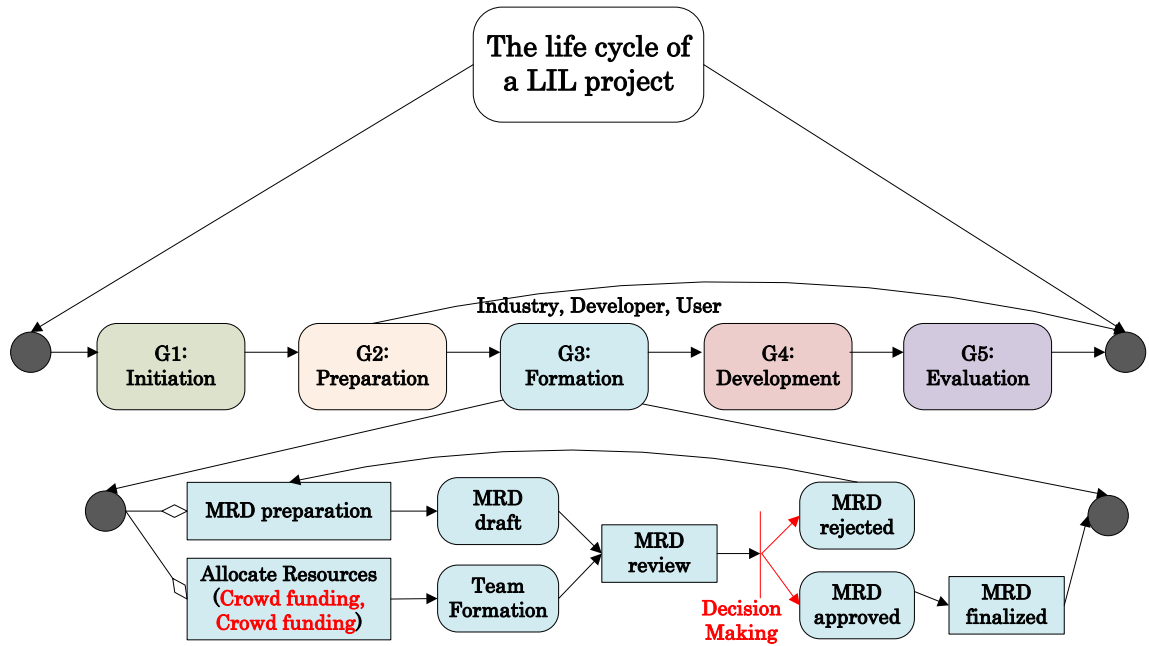

Figure 4.10: The composite goal of LIL Formation from Goal Net perspective

This phase, as the last step before actual development, is to finalize the detailed product design and make sure that the developers and testers fully understand the requirements. After the project is granted by investors (industry) or crowd funding (user), LIL needs to recruit talents and form a team working in some place, where crowd sourcing can be used sometimes. Meanwhile, market requirement document (MRD) needs to be prepared to illustrate what the product should look like and function as. It is also a useful reference in testing and evaluation section as well as possible future reviews.

**Phase 4 –** *LIL Development:*

(Involved parties: Industry, Developer, User. Keywords: crowd sourcing)

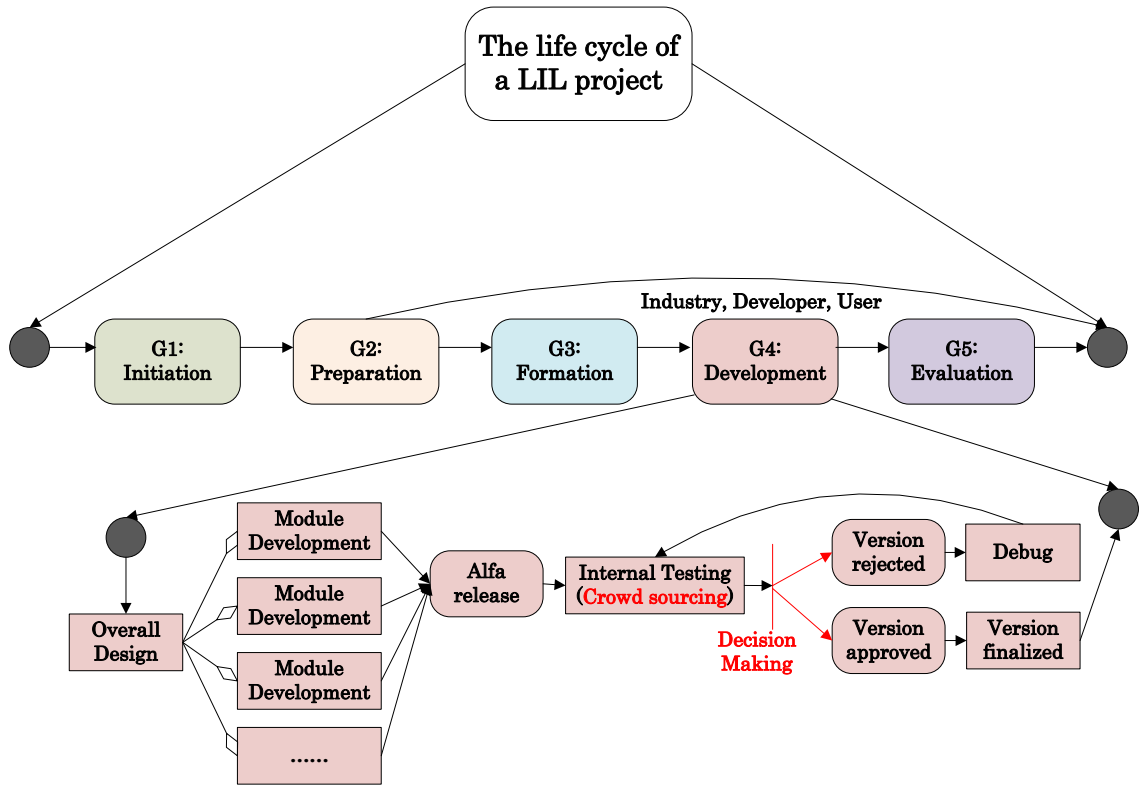

Figure 4.11: The composite goal of LIL Development from Goal Net perspective

Generally, there are three versions to be released step by step in a product life cycle: alpha, beta, and gamma. Alfa release is the first phase to begin product testing. Before the product can be released to the public, it should be tested by developers themselves with white-box techniques first and then additionally validated by other team members within the organization with black-box techniques [42]. Crowd sourcing may involve more people, like shareholders and outside co-developers, in internal testing to ensure it is bug-free enough to be released to the public.

**Phase 5 –** *LIL Evaluation*:

(Involved parties: Industry, Developer, User. Keywords: crowd sourcing)

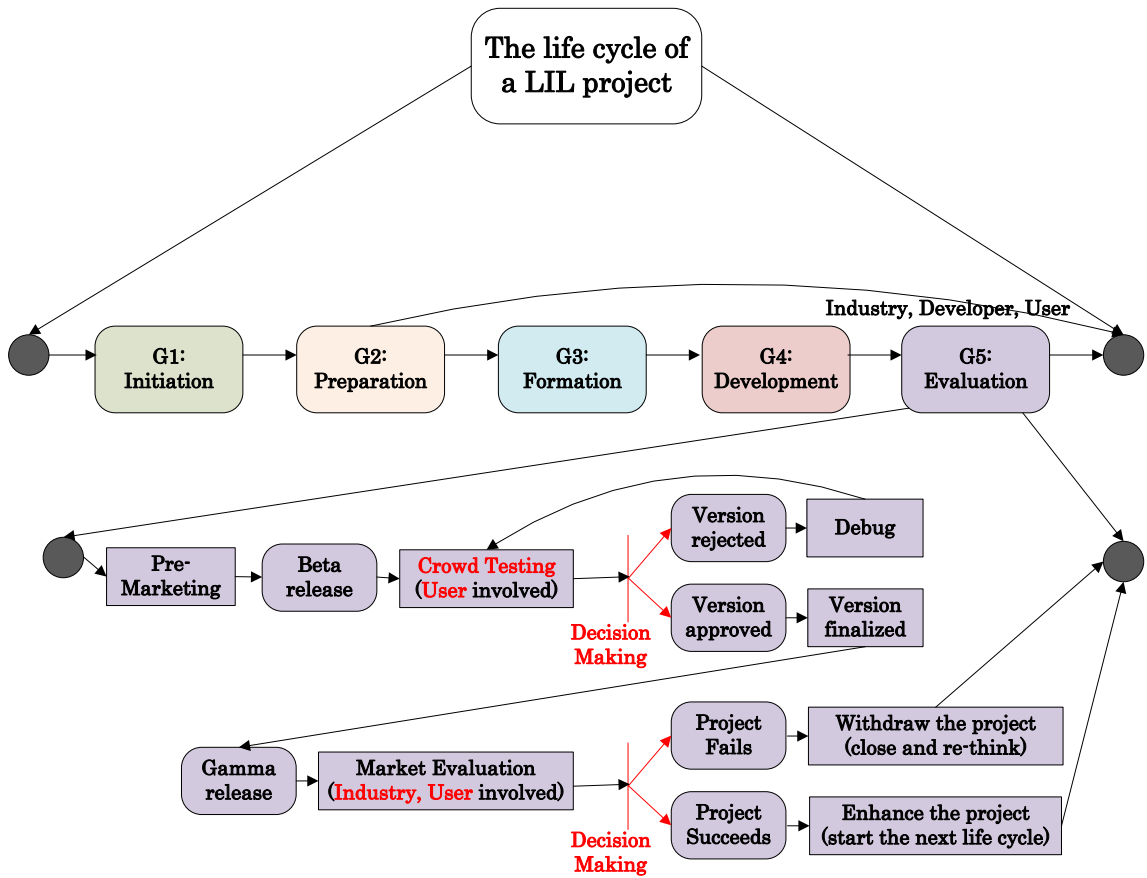

Figure 4.12: The composite goal of LIL Evaluation from Goal Net perspective

The last phase requires all parties to evaluate the project together. Beta release is a complete version for public testing. It is better to do some marketing before release to attract enough testers to download or access the beta version to perform testing and give feedbacks as normal users. Crowd testing could gather as many public users as possible for remote usability testing under diverse realistic platforms. After the revision based on the public testing feedbacks, a more stable version is officially released, so call "gamma release". The purpose of evaluation will move from usability evaluation to some realistic questions: whether it is popular enough to obtain a satisfied market share, how much revenue it can generate, whether it is worth to invest another round, and so on. The real-time market evaluation will determine how successful the product is and whether to continuously improve the product or just close the project. It is worth to mention that the failure of the project does not mean LIL fails. As long as LIL runs through the five phases with high effectiveness and efficiency, it can be marked as success no matter the product is successful or not. The timely cancellation of a failing product also can reflect the decisiveness of LIL.

## 4.3 Some highlights in the methodology

The degree of player involvement in LIL Model Design based on Goal Net is higher than other LIL models, which is summarized in the table below. The red crosses show that user and industry party did not participate in phase 2 and 3, but now get involved. As a result, all parties are fully involved throughout the life cycle of an LIL project, to achieve **MC**. Besides, advanced methods, including big data, crowd sourcing, crowd funding and crowd testing are applied in the different phases, in order to achieve **UUI** and **MC** and eventually upgrade LIL from 1.0 to 2.0.

|  | Advanced methods | Involvement | | |
|---|---|---|---|---|
|  |  | User | Developer | Industry |
| **Phase 1:** LIL Initialization | Big data Crowd souring | x | x | x |
| **Phase 2:** LIL Preparation | Big data Crowd souring | x | x | x |
| **Phase 3:** LIL Formation | Crowd funding Crowd souring | x | x | x |
| **Phase 4:** LIL Development | Crowd souring | x | x | x |
| **Phase 5:** LIL Evaluation | Crowd testing | x | x | x |

Table 4.1: The advanced methods and player involvement in Living Laboratory Model based on Goal Net

In addition, PCA is applied in each of decision making nodes by considering different factors and anticipating the future trends.

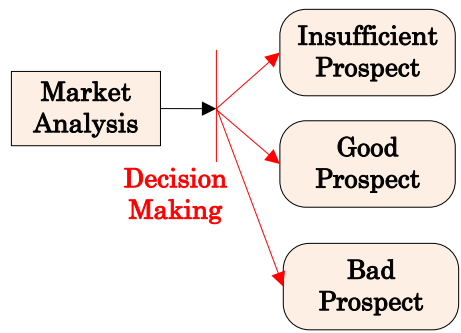

Figure 4.13: The decision making node in phase 2 LIL preparation

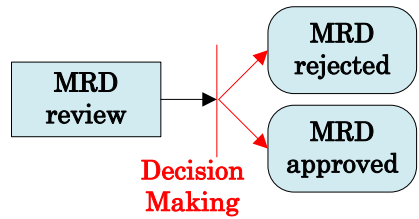

Figure 4.14: The decision making node in phase 3 LIL formation

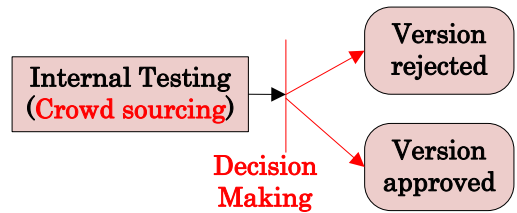

Figure 4.15: The decision making node in phase 4 LIL development

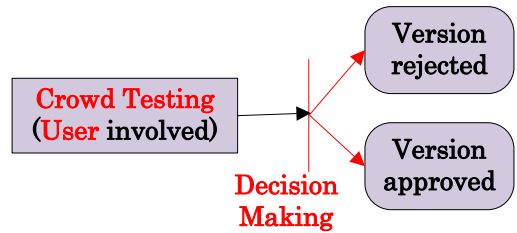

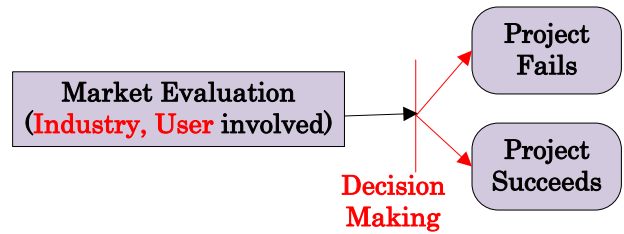

Figure 4.16: The decision making node in phase 5 LIL evaluation

In the decision making nodes above, the following factors should be considered to understand the current situation and predict the future context.

$$F_{dm} = \sum_{k=0}^{n} f(\text{cost, time, user benefits, investment returns, tradeoffs, technical constraints, ...})$$

The value of each factor can be obtained in the help of big data and other data collection methods. As the relationship and importance of the factors vary in different projects according to real situation, the detailed formula cannot be generally made. But it is a necessary step to ensure the PCA characteristics in LIL 2.0.

## 4.4 Summary and Discussion

Goal net is a composite goal model to formularize a progress in a dynamic environment. It is composed of states, transitions and environment variables. A goal is a desired state to reach. A transition represents a goal relationship between goals. The whole process is affected by the environment or so-called context. As a Goal Net process is similar with an LIL project, LIL model design based on Goal Net is proposed.

In this model, big data and crowd sourcing can inspire more creative ideas in phase 1 and help the preparation of product proposal in phase 2. Crowd sourcing also attracts more talents to join the development team in phase 3 and co-create the product in phase 4 and 5. Crowd funding does not only bring more investments in phase 3, but also more people who will participate in designing and testing the product as shareholders and potentially become the end users in phase 4 and 5. Crowd testing happens in phase 5 to ensure the usability of the product. By involving all parties in each phases of LIL, UUI and MC are well practiced. In the total of five decision making nodes in LIL, we should consider many factors to predict the context changes and make a better decision. Sometimes, big data can uncover the hidden fact and help us to anticipate the future.

By following such a detailed methodology and be aware of the context it mentions, people would be able to build LIL 2.0 even if they have no experience about LIL. Some real projects adopted this methodology and benefited a lot from LIL 2.0, which will be illustrated in the next chapter.

# Chapter 5:

# Living Innovation Laboratory Implementation based on Goal Net

Since the new generation of LIL and the detailed methodology towards it were fully described in the previous chapters, it was time to verify them in the real world projects. In the past one year, I have been involved in two projects. Both of projects got improved and even overcame their bottlenecks after applying LIL Goal Net Model. The great outcomes showed the advantages of LIL 2.0 and the helpfulness of LIL Goal Net Model.

## 5.1 Develop a Game based on Living Innovation Laboratory Goal-Net Model

The first project is used to assess the feasibility of LIL Goal Net Model. There are two research questions to be addressed: 1) whether LIL Goal Net Model can be used to build LIL successfully, 2) what difference between before and after we transformed it to an LIL project.

### 5.1.1 Context and Area of Concern

The Joint NTU-UBC Research Centre of Excellence in Active Living for the Elderly (LILY) is a world-class research centre based in Singapore, focused in promoting an active and independent lifestyle for the elderly. To address this topic, age-friendly silver games are developed in LILY.

Then, many questions come out. What kind of game should we develop? Through which device? What kind of game scenario may the elderly like? Will the elderly like this game? They are the questions we are concerned about and should keep in mind,

In order to deal with these questions well, we conducted the project in LIL Goal Net Model which could ensure the user centered design and co-creation productivity/

### 5.1.2 Methodology and Implementation

In the beginning of **phase 1**, we had interviews with domain experts, such as doctors and nurses in the elderly healthcare centers. They had a lot of experience to take care of the elderly, might know more about the elderly needs, and shared it with us. Getting inspired by the sharing, we did some data research to see what theories we had better apply in our product. We also discussed with hospitals and knew about their expectations, because they might be our potential clients. Based on the feedbacks collected from domain experts, clients, and researchers, several games with different stories and scenarios were proposed. When we cannot decide which game is the best and worth of further development, it is better to let the users choose. We set up a booth with several game prototypes in a community event, where more the two thousands of the elderly attended, and invite the **crowd** for playing. Through the **unobtrusive** observation, we found that the elderly were addicted to a Kinect game about table tennis, as it was intuitive and easy for the elderly to play.

Based on this **crowd sourcing** way, we decided the product idea. Then we proceeded to **phase 2 and 3** where we should fully design the game. We analyzed the **big data** from the log recorded in the event day, to understand which parts the elderly liked to repeat and which parts were too difficult for the elderly to play. In the game design, we should avoid the problems and enhance the popular functions. The combination of user study and theoretical design made sure the game design could eventually meet user expectations.

In **phase 4**, the game was developed and ready for internal testing before release to the public. We invited other project teams to jointly test the game. According to the **crowd testing** feedbacks, we fixed the bugs and enhanced the game's usability. Finally, we released a stable version to the public in **phase 5**. The big data from users would be collected and analyzed as follows. Those user feedbacks from diverse users and real world context could help us to further improve the **user centered** product. Also, doctors, nurses and hospitals were invited to give comments so that they could **co-create** the product.

Through the five phases, we produced a game product by applying the LIL goal-nect model. It became a LIL project which would be continuously improved based on the user centered co-creation approach.

### 5.1.3 Outcomes

To analyze the outcomes, phenomenography was chosen as a qualitative research method to study human experiences and evaluate the experiment outcomes.

#### 5.1.3.1 Introduction to Phenomenography

Phenomenography refers to "a research method for mapping the qualitatively different ways in which people experience, conceptualize, perceive, and understand various aspects of, and phenomena in, the world around them" (Marton, 2001, p.144). It was first adopted in the University of Goteborg, Sweden, in the early 1970s, to investigate why some students studied better than others. Through phenomenography, researchers looked into the content aspect of learning and the act aspect of learning. In the end, they found the different ways students understand the content of learning and the different ways students experience the learning situation and their act of learning (Marton, 1997). Now, phenomenography is a famous qualitative research method, widely used in Australia, the Netherlands, and the United Kingdom (Richardson, 1999).

Phenomenography is to study how people experience a given phenomenon, not to study a given phenomenon [45]. Marton (1981) highlighted that phenomenography "aims at description, analysis, and understanding of experiences; the **object** of the research is the **variations in ways of experiencing phenomena**" (p.180). It reveals the variations through describing the phenomena in the world as people see them (Marton & Booth, 1997, p. 111). Phenomenography studies the way people experience something, which "is an internal relationship between the experiencer and the experienced" (Marton & Booth, 1997, p.113). A phenomenographic study was demonstrated in Figure 10.2, which showed that the object of a phenomenographic study is the relations between the subjects and the phenomenon, instead of the phenomenon itself.

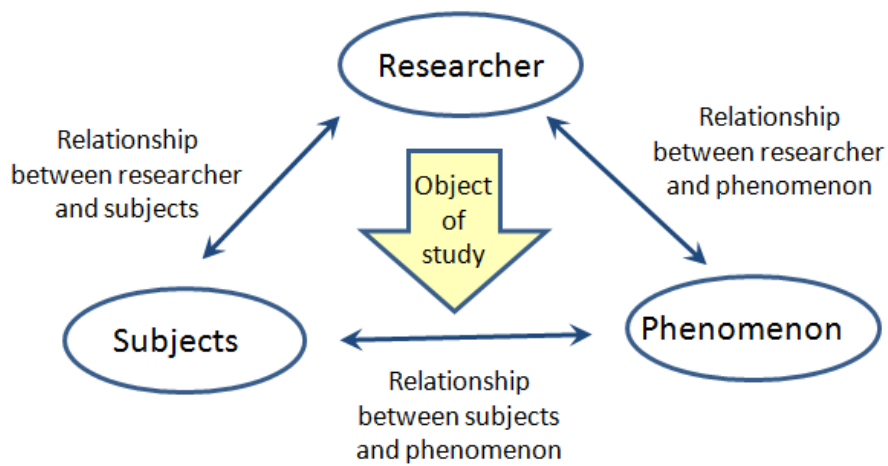

Figure 5.1: Phenomenographic Relationality (Bowden, 2005, p.13)

The **result** of phenomenographic research is a collective analysis of individual experiences. Data is collected at an individual level, but the aim is to find the collective awareness and variation in how a phenomenon is experienced (Marton & Booth, 1997, p.114). According to the principle of phenomenography, "whatever phenomenon we encounter, it is experienced in a limited number of qualitatively different ways" (Marton & Booth, 1997, P122). Through the description of the variation in ways people experience phenomena, different categories of human experiences are found and the essence of the variation is captured (Marton & Booth, 1997, p121). So, the outcomes of phenomenographic research comprise the limited **categories of experiences as well as their relations** found in variation analysis [45]. Åkerlind (2005) highlighted that the description and interpretation of variation in experience in a useful and meaningful way would reveal what would be required for individuals to move from less powerful to more powerful ways of understanding a phenomenon (p.72).

Phenomenography is **different from phenomenology** in terms of the object of research. Phenomenology researcher is exploring his or her own experience by reflecting on it, while phenomenography researcher is exploring other people's experiences by reflecting on them (Marton & Booth, 1997, P120).) Phenomenology emphasizes philosophy and psychology. It assumes that there are many ways to interpret the same experience and the meaning of the experience to each person. Phenomenology aims to describe and interpret an experience by determining the meaning of the experience as perceived by the people who have participated in it (Ary, Jacobs & Sorensen, 2010, p. 471). In addition, phenomenology tries to move from individual experience to a universal essence, in order to determine the essence of the experience as 'perceived by the participants' (p. 472). In contrast, phenomenography focuses on "investigating the experience of others and their subsequent perceptions of the phenomenon - their reflections on the phenomenon"(p.474).

### 5.1.3.2 Choose Phenomenography as a Qualitative Method

Why did we choose phenomenography as a qualitative method to evaluate the project outcomes? LIL Goal Net Model is a new way to build LIL 2.0. Before applying it to the projects in LILY lab, people in LILY lab had no knowledge and experience about LIL 2.0. Their experiences and perceptions before and after were expected to be quite different, which was perfect scenarios for phenomenographic study.

First, phenomenographic study focuses on categorizing different human experiences about a given phenomena and finding the way to empower human towards better outcomes. In this project, we tried to survey people's experiences before and after applying LIL Goal Net Model, in order to determine whether the proposed methodology was the better way to conduct an efficient and sustainable project. Second, phenomenography requires a second order perspective, which means researchers should analyze other people's perspectives about the given phenomenon, instead of making statements about the phenomenon themselves from the first order perspective [45]. To evaluate LIL Goal Net Model applied in this project, we should employ a second order perspective. We investigated the experiences of developers and users involved in the project from the second order perspective. In this way, the feedback on LIL Goal Net Model was from the participants' perspective, rather than the researcher's interpretation.

Therefore, it should be quite suitable to adopt phenomenography as a qualitative method to evaluate LIL Goal Net Model applied in this project.

### 5.1.3.3 Phenomenographic Analysis and Results

*Participants*: There were two groups of participants in the phenomenographic study, that is 12 users and 4 developers who co-created the project.

*Procedures*: Users and developers were asked to complete two different set of questionnairs, followed by interviews for around 10 minutes per person.

*Data Collection & Analysis*: The following data was collected

- *Questionairs*: a survey form was desigend to address the two research questions mentioned before. 1) Can we use LIL Goal Net Model to build LIL successfully? 2) What difference between before and after we transformed it to an LIL project? Their feedback showed that, after applying LIL Goal Net Model, 80% of them felt the project got improved, 50% of them felt user involvement became deeper, 60% of them felt the co-creation was emphazied, and 80% of them felt they were more confident to set future strategies for the project.

| After applying LIL Goal Net Model, I felt that … | % of people Agree | % of people Neutral | % of people Disagree |
|---|---|---|---|
| Overall project was improved | **80%** | 10% | 10% |
| User Involvement was improved | **50%** | 40% | 10% |
| Co-creation was improved | **60%** | 40% | 0% |
| Context aware was improved | **80%** | 20% | 0% |
| Team Spirit was improved | **80%** | 10% | 10% |

Table 5.1: The survey conducted in the project team to show the success of LIL Goal Net Model

- *Interview*: The interviews were conducted individually, in order to deeply know about their indivitual experiences and understand each person's perception without others' interference and from the second-order perspective. The main conclusion was that overall **90%** of them strongly recommended this methodology to buid LIL. During the interview, 80% of them expressed that they felt more confident and active now.

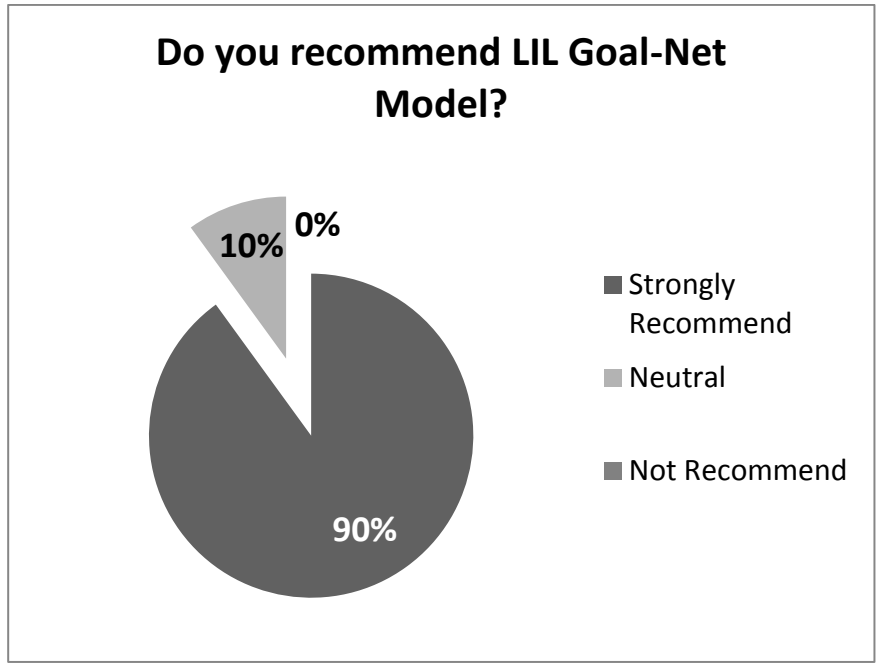

Figure 5.2: The survey conducted in the project team to show the success of LIL Goal Net Model

*Evaluation Conclusion*: This phenomenographic evaluation involves the investigation of people's experiences and perceptions in the project. Structure survey form was used to describe people's experiences; unstructured interviews was conducted to gather people's perceptions. The data collected from different participant groups were slightly different but almost indicated the same conclusion. The data in Table 5.1 and Figure 5.2 showed that, LIL Goal Net Model did not only guide us to build an LIL project, but also made the team more active and productive. The main three charateristics of LIL (user centered, co-cration, and context aware) were emphasized and enhanced. Now, LIL Goal Net Model is still used in LILY research centre, in the hope of developing more LIL projects.

## 5.2 Develop an Internet Product based on Living Innovation Laboratory Goal-Net Model

Through the development of this project, we would see whether LIL Goal Net Model is able to improve a project towards LIL 2.0. There are two research questions to be addressed: 1) whether LIL Goal Net Model can transform a project to LIL 2.0, 2) what difference between before and after we transformed it to an LIL 2.0 project.

### 5.2.1 Context and Area of Concern

Baidu is the most commonly used Chinese search engine in the world with its superior search technology. In order to provide the best user experience, we designed specific search results for certain search keywords. When users search these keywords, they can easily find what they want. For example, if users search a disease name, we assemble and categorize its related knowledge into six groups "treatment, symptom, cause, diet therapy, prevention, diagnosis" and display four of them in a reasonable sequence.

Figure 5.3: Disease related Knowledge Graph (version 1)

This special type of pre-designed search result was named as "knowledge graph (KG). It is an Internet product to improve the search result page of a search engine with semantic-search information gathered from a wide variety of sources. It provides structured and detailed information about the topic in addition to a list of links to other sites [43]. The result is that users do not need to navigate to other sites, but can quickly resolve their query by simply referring to KG. KG's quality is evaluated in terms of the CTR (click-through-ratio) of KG and the whole search result page. The higher CTR is, the higher quality of the product is. CTR is the number of times click-through (a click is made on the site links) divided by the total impressions (the number of times an advertisement was shown), expressed as a percentage [44].

$$CTR = \frac{Clicks}{Impressions} \times 100$$

For disease related KG in Figure 5.2, the challenge was that which four of the six groups of information should be chosen to display and displayed in what sequence. Definitely, we should choose the groups of information the users were most concerned about and interested in reading more, so that the users would click KG and then CTR of KG and the whole page would be increased. In the beginning, we used LIL 1.0 approach to design the product. Based on user study conducted by survey and interview, we determined to set the four and the sequence as ""treatment, cause, symptom, diet therapy", just like Figure 5.2. After releasing this KG online, it did not increase the CTR of the whole page. The CTR of KG itself was lower than expected. In order to improve the product, LIL Goal Net Model was used to develop KG as an LIL 2.0 product.

### 5.2.2 Methodology and Implementation

In **phase 1**, besides the traditional user study methods, **big data** from user sessions was used to find what information users most expected to see. A user session means when a user starts to search a keyword, which links he or she chooses to click in search result page until finding the best result he or she actually needs. The session ends only when the user stops searching anything. During the session, what kinds of websites the user visited means what kinds of knowledge he or she was interested in. For example, a user searched the keyword "diabetes" and clicked a website link titled as "How to treat diabetes". That means he or she might be interested in the knowledge about treatment. If By analyzing the big data from 31193 user sessions, it was found that the links users most often clicked were the site about "general information (XX times), treatment (XX times), symptom (XX times), diet therapy (XX times)". But no one had suggested "general information" before.

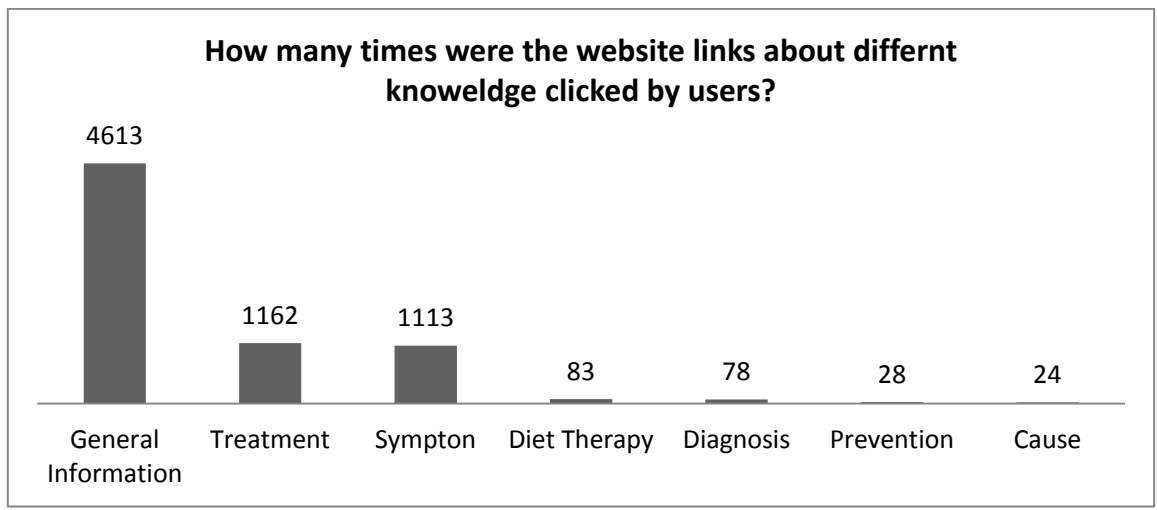

Figure 5.4: Disease related Knowledge Graph (version 2)

The data showed what kinds of knowledge the users are most interested in and tended to click. We should display them to the users at the first place. The top one was "general information" and the last one was "cause". Even users did not notice and suggest it. **UUI** helped us to discover the hidden truth. So we changed the contents of KG to the following one:

Figure 5.5: Disease related Knowledge Graph (version 2)

**Crowd testing** was applied in **phase 5** to achieve **MCC** in LIL 2.0. Both of new and old versions of KG were released online and randomly displayed to different users. Users had 20% possibility to see the new version and 20% possibility to see the old version. The two versions had the almost same sample users, but they got different performance in terms of CTR. Then, the better version was found according to user selection.

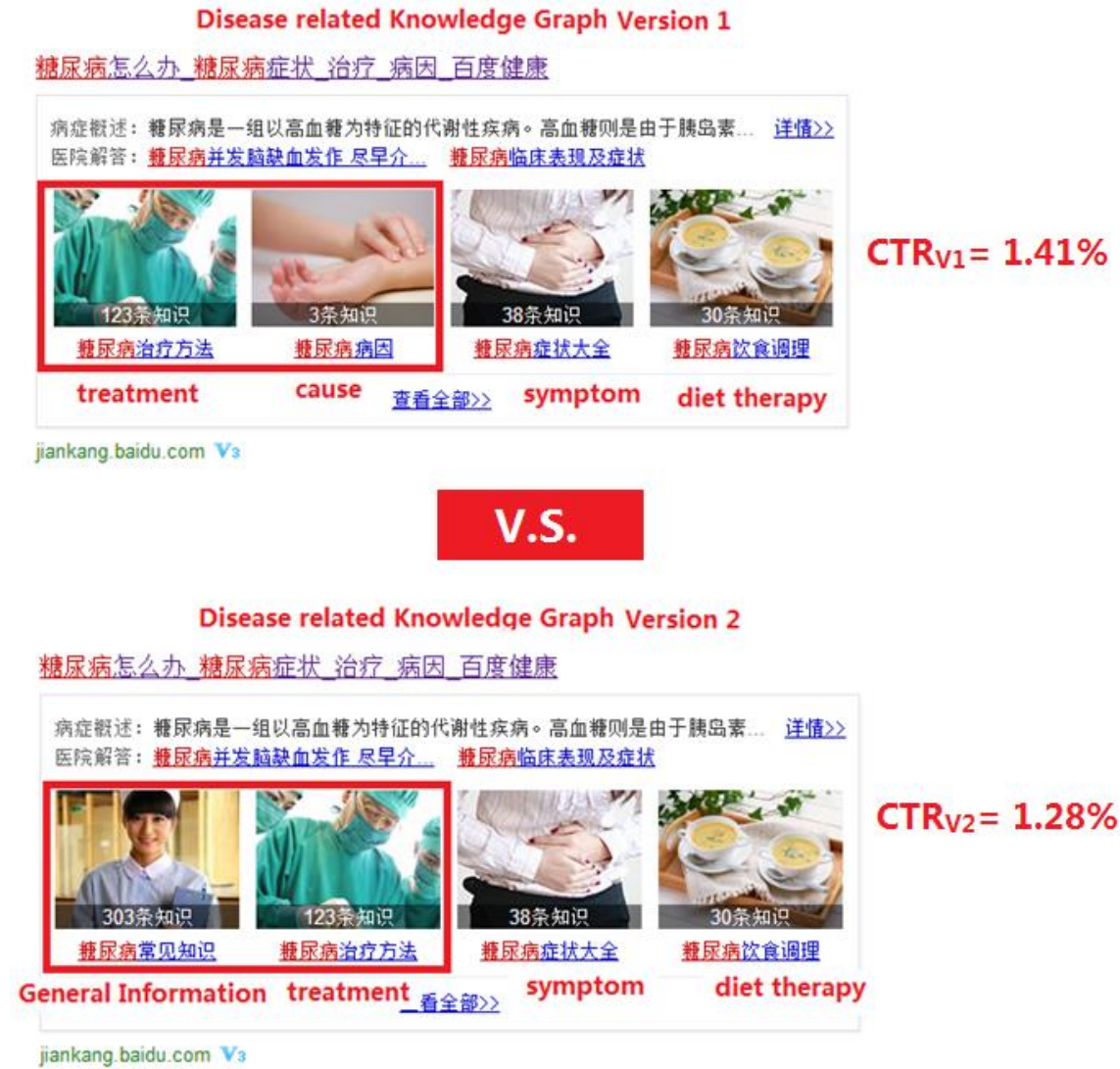

Figure 5.6: The CTR comparison of two versions of Knowledge Graph

In the end of 2013, we needed to set strategies for the coming 2014. So we looked into big data and found the medical market share in PC kept dropping in 2013. We anticipated that the medical users would move from PC to mobile.

Hence, we set our main strategy for 2014 was to make a lot of efforts on mobile applications.

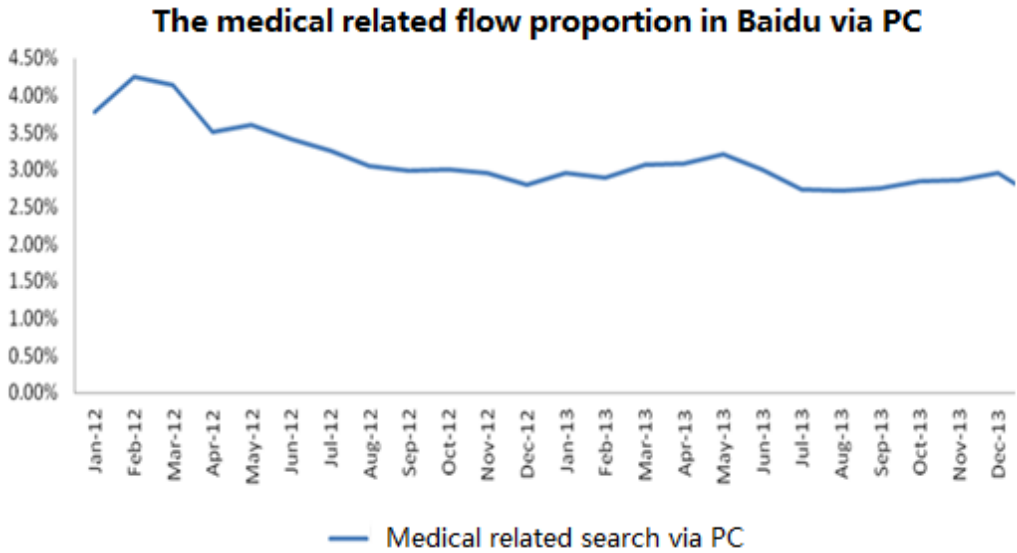

Figure 5.7: PC market became smaller in 2013, based on big data analysis.

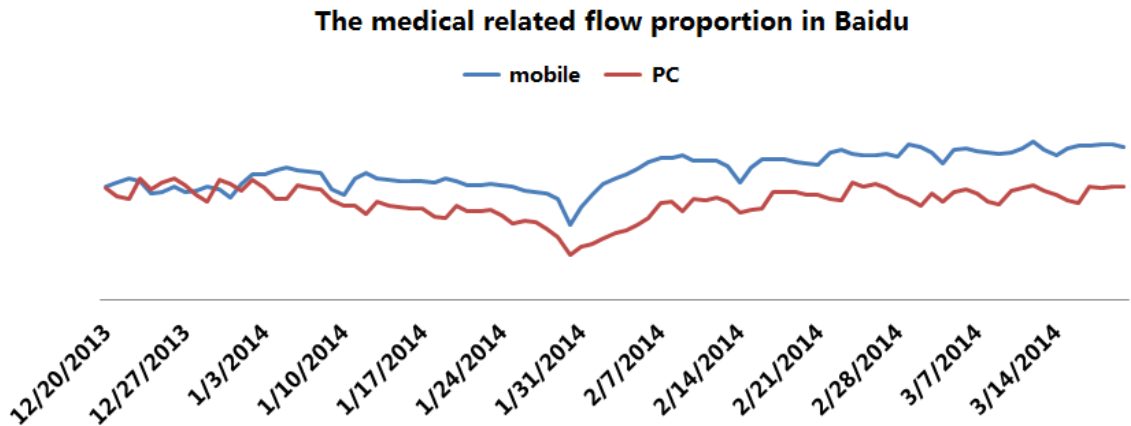

Figure 5.8: Mobile market overcame PC market in 2014

The truth is that mobile market overcame PC market in 2014. Luckily, we have already prepared a lot to capture the expanding mobile market. PCA guided us to forecast the potential change in the future, so that our products could survive for a long time.

### 5.2.3 Outcomes

Phenomenography was chosen as a qualitative research method to evaluate the project outcomes.

*Participants*: Search engine users were divided into two groups: one group of them could only see the old version of KG, while the other group could only see the new version.

*Procedures*: The big data of user behaviors were collected and analyzed to reflect user perference. The change of user experiences before and after applying LIL Goal Net Model would evalute the success of this methodology.

*Data Collection & Analysis*: The following data was collected and analyzed, to address the two research questions mentioned before. 1) Is LIL Goal Net Model able to transform a project to LIL 2.0? 2) What difference between before and after we transformed it to an LIL 2.0 project?

- *Big data*: As LIL Goal Net Model guided us to use big data and crowd testing in right phases, we made the product as an LIL 2.0 product with the characteristics of UUI and MCC. Within 8 days, the big data of 475,454 user behaviors were tracked and recorded for phenomenographic analysis (237,511 user behaviors for the old version; 237,943 user behaviors for the new version). The big data of user behaviors showed that the quality of the product was increased in terms of CTR of KG (+0.71%) and the whole page (+10.17%).

|  | Version 1 | Version 2 | Comparision |
|---|---|---|---|
| CTR of KG | 8.10% | 8.16% | + 0.71% |
| CTR of page | 116.57% | 128.43% | + 10.17% |

Table 5.2: The CTR comparison of two versions of Knowledge Graph

*Evaluation Conclusion*: This phenomenographic evaluation involves the investigation of people's experiences and perceptions in the project. The big data of user behaviors did not only describe user experiences, but also represent user prepferences. Table 5.2 showed that the product broke its bottleneck after transferring to an LIL 2.0 product. **The CTR of KG increased by 10.17% relatively**; the CTR of the whole page increased by 0.71% relatively. It was a very significant improvement already. It meant that people preferred the new version designed derived from LIL Goal Net Model. So, the two research questions could be answered by saying that LIL Goal Net Model did not only transform the disease related KG to an LIL 2.0 product, but also made its quality improved a lot.

## 5.3 Summary and Discussion

In order to validate the concepts of LIL 2.0 and LIL Goal Net Model, it has been applied to two real world projects. Phenomenography is a qualitative research method to study human experience and their relations. It was used in our research to evaluate the project outcomes after applying LIL Goal Net Model to them.

First, a Kinect game for the elderly was built in LIL Goal Net Model. It involved a wide range of users to participate in design phase, by using big data. Many users were also invited to development and testing process as co-creators. The two key concepts UUI and MCC were well established and realized in the life cycle of this project. In phenomenographic study, the project team, including users and developers, all commented that the quality and productivity of the project were both improved.

The second case is Baidu's KG product, which I have participated in the project for over one and a half years. The product got significant improvement after it was developed in the concept of LIL 2.0. In the beginning, we could not find the real requirements of users through the traditional user study method in LIL 1.0. Luckily, UUI helped us to discover it through big data. Based on it, we designed a new version of KG. In phenomenographic study, big data technology was used. The big data of user behaviors showed that the CTR of the whole page was increased significantly by 10.17% after applying LIL Goal Net Model.

Through the two projects and their corresponding phenomenographic studies, the conclusions could be drawn as follows:

- LIL can be well implemented by using LIL Goal Net Model.

- LIL Goal Net Model can transform a project to LIL 2.0.

- The project should be improved after upgrading to LIL 2.0.

# Chapter 6:

# Conclusion and Future Recommendation

This chapter summarized the research works done in this paper, to answer the two research questions mentioned at the beginning. There are also some questions and improvements for us to think and explore in the future.

## 6.1 Conclusion

LIL refers to a research and development approach where innovations, such as products or services, are co-created and validated in collaborative and multi-contextual real-world environments [3].

With the aim of developing innovative solutions based on users' needs, the implementation should be based on user involvement throughout the innovation process, thereby making LIL **user centered**, as opposed to technology centric [3]. Besides, he success of LIL depends on how well its developers, users, and industry party can cooperate together and **co-create** an innovation in a highly human interactive environment. Lastly, as the world is changing so fast now, any innovative idea may be out of day in any minute or fall into Red Ocean with tons of competitors. So LIL also requires us to conduct a project with the mindset of **context aware**. To sum up, LIL has three main characteristics: user centered, co-creation, and context aware, which makes it distinguished from other innovation approaches.

Generally, the life cycle of an LIL project is consisted of five phases: Phase 1 – Initialization, Phase 2 – Preparation, Phase 3 – Formation, Phase 4 – Development, Phase 5 – Evaluation. The roles involved in LIL include developers, users and industry party (such as shareholders, investors, sponsors, clients, and so on). They jointly develop a user centered product in a collaborative and multi-contextual environment.

After reviewing the existing concept of LIL in chapter 2, there were two research questions arising and then answered in the rest chapters.

RQ 1: *Can we refine the three characteristics of LIL? Can we define a new generation of LIL with the new characteristics?*

As the existing LIL tend to generate incremental innovation, it should be refined to LIL 2.0 in order to generate disruptive innovation. LIL 2.0 has three characteristics of UUI, MCC, and PCA. UUI means users can contribute to the project in multiple ways. User data can be even collected in an unobtrusive way, such as big data. UUI helps the development team to discover user needs which they may not even realize themselves. MCC is to gather a large number of people (including users and industry party) with diverse and random backgrounds to co-create a project. Crowd sourcing, crowd testing, and crowd testing can be used to achieve MCC. PCA means that we should forecast the potential changes of the context through big data and update our strategies accordingly to prevent risks in the future. With these three newly defined characteristics, the concept of LIL 2.0 emerges with more effectiveness, efficiency and lower business risk. How LIL 2.0 is advanced than LIL 1.0 was illustrated in chapter 5. Two projects (a Kinect game and an Internet product) got improved after becoming LIL 2.0 projects and finally broke their bottlenecks.

RQ 2: *Can we have a detailed methodology to build this new generation of LIL?*

LIL Goal Net Model is a detailed and systemic methodology for LIL 2.0, based on the theory of Goal Net Model. By following the methodology, any person could build an LIL 2.0 project even without experience about LIL. All advanced methods, including big data, crowding sourcing, crowd funding, and crowd testing, are adopted to achieve UUI, MCC and PCA in LIL 2.0. This methodology was applied to two real world projects to prove its feasibility. In the first project, a Kinect game was co-created by users, developers and industry party and the feedbacks from team members showed the success of applying LIL Goal Net Model to build LIL 2.0. The second project is Baidu's KG. Following LIL Goal Net Model, the project changed to be LIL 2.0. The quality of the product was improved, measure by its CTR.

In conclusion, LIL 1.0 gathers a group of people to co-create a user centered innovation in the changing context, while LIL 2.0 gathers a large number of diverse people to join development and realize an innovation idea derived from obtrusive and unobtrusive user study. LIL 2.0 is advanced than LIL 1.0 in terms of the degree of user involvement, the productivity of co-creation, and business risks. LIL Goal Net Model is a useful methodology to build LIL 2.0, which has been proved in the implementation of two real projects.

## 6.2 Future Recommendation

During the implementation of LIL projects, some aspects were identified which we believe are important to do more research about. These aspects are related to the threshold of big data and scammer issue in crowd testing.

First, although there are countless data in the cyber world, not every product is able to accumulate big data within a short time. It depends on how large the potential market is, hardware limitation, how famous the brand is, and many other variables. For example, an unknown website takes a lot of time to accumulate big data, while Alibaba can easily reach billions of users and their shopping behavior data within one day. Besides, it must consume a lot of time and efforts on big data analysis. Big data refers to large data sets with various formats and qualities. The first step of big data analysis is to clean, enrich, and harmonize the data. The big data analyzers in Silicon Village of USA usually spend 50-80% of efforts on data cleaning and then proceed to data analysis, where more data mining technologies should be used. Not all of companies and organizations have enough resources to conduct big data analysis. If the data is not clean enough or handled unprofessionally, the result of big data analysis may totally different and even inverse. Hence, the threshold of big data is too high, based on the current technology. More research on big data should be conducted in the future. Otherwise, it is hard to used big data to achieve UUI and PCA in LIL 2.0.

Second, crowd testing may be abused by scammers. In crowd testing, there are 2 types of testers: serious workers & scammers. Scammers just guess the answers without reading the question. They have 50% chance to guess out the correct answer and get payment. Scammers can earn more rewards than serious workers within the same time. Amazon Mechanical Turk is a crowd testing platform. However, due to the large number of scammers, 40.92% of testing tasks are completed but totally useless. Test requesters need to spend more money and manpower to clean the data and sort out the correct results. It results in high cost and risk, which many LILs cannot afford.

Therefore, more research is expected to be done on big data and crowd testing. Nowadays, there are more and more papers about big data technology published. It should be possible to lower its threshold in the near future. For crowd testing, trust model and active learning strategies should be helpful to prevent scammers. If the two issues can be solved, big data and crowd testing could be widely used in LIL.

# Appendix

## A.1 Survey form to evaluate the LILY project after applying LIL 2.0 and LIL Goal Net Model

The degree of user involvement, co-creation and context awareness indicates the thoroughness of an LIL project. Table A.1 showed the key points to evaluate an LIL project. The survey questions were designed around these key points. One survey form was designed for developers and the other one was designed for users to fill up.

| After applying LIL Goal Net Model, I felt that … | % of people Agree | % of people Neutral | % of people Disagree |
|---|---|---|---|
| Overall project was improved | | | |
| User Involvement was improved | | | |
| Co-creation was improved | | | |
| Context aware was improved | | | |

| Team Spirit was improved | | | |
|---|---|---|---|

Table A.1: The key points to evaluate an LIL project [6]

---

**After applying LIL goal net methodology to the project:**

1. How is the project now, compared with previously?

   ☐ Improved  ☐ The same  ☐ Worse

2. What do you feel about the degree of user involvement in the project?

   ☐ Deeper  ☐ The same  ☐ Shallow

3. How is the co-creation in the project now?

   ☐ More efficient  ☐ The same  ☐ Stumbled

4. How is your context awareness now?

   ☐ Acuminous  ☐ The same  ☐ Slow

5. How do you feel about your project now?

   ☐ More confident  ☐ The same  ☐ More anxious

6. Do you recommend these methods to the future projects you will be involved?

   ☐ Yes  ☐ Neutral  ☐ No

Table A.2: The survey form for developers to evaluate an LIL project

**After applying LIL goal net methodology to the project:**

1. How do you feel about the project now, compared with previously?

    ☐ Improved ☐ The same ☐ Worse

2. What do you feel about your involvement in the project?

    ☐ Deeper ☐ The same ☐ Shallow

3. How is your participation in project co-creation process now?

    ☐ Deeper ☐ The same ☐ Shallow

4. How is your context awareness now?

    ☐ Acuminous ☐ The same ☐ Slow

5. How is your feeling now as a user participant in the project?

    ☐ More self-worth ☐ The same ☐ More anxious

6. Do you think these methods should be applied to other projects?

    ☐ Yes ☐ Neutral ☐ No

Table A.3: The survey form for users to evaluate an LIL project

## A.2 Big data of user behaviors to evaluate the KG project after applying LIL 2.0 and LIL Goal Net Model

Within 8 days, the big data of 475,454 user behaviors were tracked and recorded for phenomenographic analysis (237,511 user behaviors for the old version; 237,943 user behaviors for the new version). The big data of user behaviors showed that the quality of the product was increased in terms of CTR of KG (+0.71%) and the whole page (+10.17%).

|       | User behaviors | | CTR of KG | |
|---|---|---|---|---|
|       | version 1 | version 2 | version 1 | version 2 |
| Day 1 | 25426 | 25417 | 7.85% | 8.12% |
| Day 2 | 25113 | 25105 | 8.57% | 7.14% |
| Day 3 | 33866 | 33873 | 8.16% | 8.58% |
| Day 4 | 28903 | 29023 | 8.14% | 7.22% |
| Day 5 | 26978 | 27101 | 8.10% | 8.23% |
| Day 6 | 27315 | 27298 | 7.95% | 8.84% |
| Day 7 | 36274 | 36332 | 7.79% | 8.96% |
| Day 8 | 33636 | 33794 | 8.27% | 8.22% |
| **Total** | 237511 | 237943 | **8.10%** | **8.16%** |

Table A.4: The big data of user behaviors to show the increase of CTR of KG

|  | User behaviors | | CTR of whole page | |
|---|---|---|---|---|
|  | version 1 | version 2 | version 1 | version 2 |
| Day 1 | 25426 | 25417 | 117.01% | 127.15% |
| Day 2 | 25113 | 25105 | 116.55% | 128.28% |
| Day 3 | 33866 | 33873 | 116.69% | 127.52% |
| Day 4 | 28903 | 29023 | 117.46% | 125.69% |
| Day 5 | 26978 | 27101 | 114.72% | 126.85% |
| Day 6 | 27315 | 27298 | 117.85% | 131.69% |
| Day 7 | 36274 | 36332 | 116.92% | 129.42% |
| Day 8 | 33636 | 33794 | 115.35% | 130.85% |
| **Total** | 237511 | 237943 | **116.57%** | **128.43%** |

Table A.5: The big data of user behaviors to show the increase of CTR of the whole page

## A.3  A Network of Living Innovation Laboratories

In the real contexts of living and working, any research laboratory should take note of regional policies. The concept of a network of LIL overcomes the territorial limitation. A broader territorial innovation may take place in the aim of giving benefits for the surrounding community and even beyond it.

In the past few years, regionally-oriented LIL have emerged bottom-up from the local level where development funding is easily managed and concrete benefits directly address the local communities. European Network of Living Labs (ENoLL) can meet the challenge of constructing this bottom-up structure on the European level, through:

- Regional LILs are mapping together with dynamic links based on the territorial contexts of themselves and each other;
- Specific case instances as evidences of a LIL actually giving benefits to its surrounding region;
- The local governance structures (for example, Local and Coastal Action Groups, Territorial Pacts, River and Landscape Contracts, and so on) are integrated with LILs.
- The operational and procedural integration of LILs at the local and regional level into a sustainable network at the EU level, with some related strategies and priorities, have its objectives as follows:
  o To introduce a shared community concept which assembles LILs with a regional and territorial dimension together;
  o To develop best methods to link LILs with local sustainable development objectives;

- To promote LILs concept as a development guideline at the regional, national and European levels.

European Network of Living Labs (ENoLL) is the first network of Living Innovation Laboratories in the world. Since November 2006, there are totally 212 LILs in ENoLL, including also 25 affiliated LILs in non European Countries.

Meanwhile, all branches of ENoLL are open for new organizations to discuss about partnership extension. As end-users, individual citizens may join the site-specific end-user community.

For organizations (such as companies, universities, regional authorities or other government organizations), partnership may be established if they aim to get engaged in LIL development and keep sustainable in operations.

As we can see, the network of LIL does not only emphasize the relationship among those LILs, but also extend to the partnership with companies, citizens, government and other societal or technological organizations.

Their essential vision is still to re-define the user communities from just stakeholders and consumers of industry to actually contributors and co-developers of new innovations.

# References

[1] Andrew Kusiak. Innovation: The Living Laboratory Perspective. *Computer-Aided Design & Applications, Vol. 4, No. 6, 2007, pp 863-876.*

[2] Esteve Almirall, Jonathan Wareham. Living Labs and Open Innovation: Roles and Applicability. *The Electronic Journal for Virtual Organizations and Networks, Volume 10, "Special Issue on Living Labs", August 2008.*

[3] Mats Eriksson, Veli-Pekka Niitamo, Seija Kulkki. State-of-the-art in utilizing Living Labs approach to user centric ICT innovation - a European approach. Centre for Distance-spanning Technology (CDT) at Luleå University of Technology, Sweden.

[4] Per Levén, Jonny Holmström. Consumer co-creation and the ecology of innovation: A living lab approach. Department of Informatics, Umeå University, Sweden.

[5]   P. Ballon, J. Pierson, S. Delaere, "Test and Experimentation Platforms for Broadband Innovation: Examining European Practice".

[6]   Tingan Tang, Combining User and Context: Living Labs Innovation in Digital Servies. Aalto University publication series, Doctoral Dissertations 91/2014.

[7]   Anna Ståhlbröst, Forming Future IT - The Living Lab Way of User Involvement. Department of Business Administration and Social Sciences, Division of Informatics, Luleå University of Technology, 12/2008..

[8]   Prahalad, C.K.; Ramaswamy, V. (2004) "Co-Creation Experiences: The Next Practice in Value Creation". *Journal of Interactive Marketing. Volume 18, Number 3*.

[9]   B.C.W. Smeets, Crowdsourcing: the process of innovation. Tilburg University, August, 2011.

[10]  "Oxford Dictionary Definition of Crowdfunding". Retrieved July 23, 2014. The Merriam Webster Dictionary defines Crowdfunding as "the practice of soliciting financial contributions from a large number of people especially from the online community" "Merriam Webster Dictionary Definition of Crowdfunding". Retrieved July 23, 2014.

[11]  Daniel Broderick, Crowdfunding's Untapped Potential In Emerging Markets. *Forbes Brand Voice, 08/05/2014.*

[12] Steinhauser, Markus, "Two Approaches to Crowdsourced Software Testing". *Retrieved 13 November 2013.*

[13] PJ Stappers, H. van Rijn, SC Kistemaker, AE Hennink, and F. Sleeswijk Visser. Designing for other people's strengths and motivations: Three cases using context, visions, and experiential prototypes. *Advanced Engineering Informatics*, 23(2):174–183, 2009.

[14] Ibrahim Abaker Targio Hashem, Ibrar Yaqoob, Nor Badrul Anuar, Salimah Mokhtar, Abdullah Gani, Samee Ullah Khan, The rise of "big data" on cloud computing: Review and open research issues, *Information Systems, Volume 47, January 2015, Pages 98-115, ISSN 0306-4379.*

[15] Chesbrough, H., Vanhaverbeke, W. and West, J. (2006), *Open Innovation: Researching a New Paradigm*, edn. Oxford: Oxford University Press.

[16] Dodgson, M., Gann, D. and Salter, A. (2006) *The role of technology in the shift towards open innovation: The case of Procter&Gamble. R&D managemt* 36, 333-346.

[17] Dodgson, M., Gann, D. and Salter, A. (2005) *Think, Play, Do: Technology, innovation, and organization*, edn. Oxford: Oxford University Press.

[18] Edquist, C. (2004) *Reflections on the system of innovation approach. Science and Public Policy* 31, 485-489.

[19] Franke, N. and Shah, S. (2003) *How communities support innovative activities: an exploration of assistance and sharing among end-users. Reseach Policy* 32, 157-178.

[20] Galanakis K. Innovation process: Make sense using systems thinking, Technovation, 26(11), 2006, 1222- 1232.

[21] Hargadon, A. and Sutton, R. (1997) *Technology Brokering and Innovation in a Product Development Firm. Administrative Science Quarterly* 42, 716-749.

[22] Henkel, J. (2006) *Selective revealing in open innovation processes: The case of embedded Linux. Reseach Pollicy* 35, 953-969.

[23] Howells, J. (1999) *Research and technology outsourcing and innovation systems: an exploratory analysis. Industry and Innovation* 6, 111-129.

[24] Kline S.J. and Rosenberg N. An overview of innovation, in: Landau, R., Rosenberg, N. (Eds.), The Positive Sum Strategy, National Academy Press, Washington, DC, 1986, 275-305.

[25] Kusiak A. and Tang, C.Y. Innovation in a requirement life-cycle framework, Proceedings of the 5th International Symposium on Intelligent Manufacturing Systems, IMS'2006, Sakarya University, Sakarya, Turkey, 2006, 61-67.

[26] Lettl, C., Herstatt, C. and Gemuenden, H.G. (2006) *Users' contributions to radical innovation: evidence from four cases in the field of medical equipment technology. R&D Management* 36, 251-272.

[27] Malerba, F., Nelson, R.R., orsenigo, L. and Winter, S.G. (2007) *Demand, innovation, and the dynamics of market structure: The role of experimental users and diverse preferences. Journal of Evolutionary Economics* 17, 371-399.

[28] Pilorget, L. (1993) *Innovation consultancy services in the European community. International journal of technology management* 8, 687-696.

[29] Prügl, R. and Schreier, M. (2006) *Learning from leading-edge customers at The Sims: opening up the innovation process using toolkits. R&D Management* 36, 237-250.

[30] S. Zhiqi. *Goal-Oriented Modeling For Intelligent Agents and their applications*. PhD thesis, Nanyang Technological University, 2003.

[31] Urban, G.L. and Von Hippel, E. (1988) *Lead user analysis for the development of new industrial products. Management Science* 34, 569-582.

[32] Utterback J. M., Abernathy W.J. (1975). A dynamic model of process and product innovation, OMEGA, 3(6), 1975, 639-656.

[33] Von Hippel, E. (2005) *Democratizing Innovation, edn. Cambridge, Massachusetts: The MIT Press.*

[34] Von Hippel, E. (1976) *The dominant role of users in the scientific instrument innovation process. Research Policy* 5, 212-239.

[35]     Von Hippel, E. (1988) *The sources of innovation, edn. New York: Oxford University Press.*

[36]     Von Krogh, G. and Von Hippel, E. (2006) *The promise of research on open source software. Management Science* 52, 975-983.

[37]     West, J. (2003) *How open is open enough? Melding the proprietary and open source strategies. Research Policy* 32, 1259-1285.

[38]     West, J. and Gallagher, S. (2006) *Challenges of Open Innovation: The paradox of firm investment in Open-source software. R&D Management* 36, 319-331.

[39]     West, J., Vanhaverbeke, W. and Chesbrough, H. (2006) Open Innovation: A research Agenda. In: Chesbrough, H., Vanhaverbeke, W. and West, J., (Eds.) *Open Innovation: Researching a New Paradigm, Oxford: Oxford University Press.*

[40]     Yundong Cai, Chunyan Miao, Ah-Hwee Tan, and Zhiqi Shen. Fuzzy Cognitive Goal Net for Interactive Storytelling Plot Design, Nanyang Technological University.

[41]     Z. Shen, C. Miao, X. Tao, and R. Gay. Goal oriented modeling for intelligent software agents. In *Intelligent Agent Technology, 2004. (IAT 2004). Proceedings. IEEE/WIC/ACM International Conference on*, pages 540{543, 2004. TY - CONF.

[42]     *"Encyclopedia definition of alpha version". PC Magazine.* Retrieved 2011-01-12.

[43]    Waters, Richard (May 16, 2012). "Google To Unveil Search Results Overhaul". *Financial Times.* Retrieved May 16, 2012.

[44]    Farris, Paul W.; Neil T. Bendle; Phillip E. Pfeifer; David J. Reibstein (2010). *Marketing Metrics: The Definitive Guide to Measuring Marketing Performance.* Upper Saddle River, New Jersey: Pearson Education, Inc. ISBN 0-13-705829-2. The Marketing Accountability Standards Board (MASB) endorses the definitions, purposes, and constructs of classes of measures that appear in *Marketing Metrics as part of its ongoing Common Language in Marketing Project.*

[45]    Xuehong Tao. Argumentative Learning with Intelligent Agents。*College of Education, Victoria University, Melbourne, Australia*, March 2014.

[46]    Marton, F. & Booth, S. (1997). *Learning and awareness*. Mahwah, New Jersey: Lawrence Erlbaum Associates, Inc. Publishers.

[47]    Marton, F. (1981). Phenomenography - Describing conceptions of the world around us. *Instructional Science*, 10, 177-200.

[48]    Marton, F. (1997). Phenomenography. In Keeves, J. P. (Ed.), *Educational research, methodology, and measurement: An international handbook* (2nd ed.) (pp. 95-101). Oxford: Pergamon.

[49]    Marton, F. (2001). Phenomenography: A research approach to investigating different understandings of reality. In R. R. Sherman &

R. B. Webb (Eds.), *Qualitative research in education: Focus & methods* (pp.141-161). London: RoutledgeFalmer.

[50] Cai, Y., Shen, Z., Liu, S., Yu, H., Han, X., Ji, J., Miao, C. (2014). *An Agent-based Game for the Predictive Diagnosis of Parkinson's Disease.* Paper presented at the 13th International Conference on Autonomous Agents and Multi-Agent Systems (AAMAS'14).

[51] Li, B., Yu, H., Shen, Z., Cui, L., & Lesser, V. R. (2015). *An Evolutionary Framework for Multi-Agent Organizations.* Paper presented at the 2015 IEEE/WIC/ACM International Joint Conference on Web Intelligence and Intelligent Agent Technology (WI-IAT'15).

[52] Li, B., Yu, H., Shen, Z., & Miao, C. (2009). *Evolutionary organizational search.* Paper presented at the Proceedings of The 8th International Conference on Autonomous Agents and Multiagent Systems-Volume 2.

[53] Lin, H., Hou, J., Yu, H., Shen, Z., & Miao, C. (2015). *An Agent-based Game Platform for Exercising People's Prospective Memory.* Paper presented at the 2015 IEEE/WIC/ACM International Joint Conference on Web Intelligence and Intelligent Agent Technology (WI-IAT'15).

[54] Lin, J., Miao, C., & Yu, H. (2011). A cloud and agent based architecture design for an educational mobile SNS game *Edutainment Technologies. Educational Games and Virtual Reality/Augmented Reality Applications* (pp. 212-219): Springer Berlin Heidelberg.

[55] Lin, J., Yu, H., Miao, C., & Shen, Z. (2015). *An Affective Agent for*

*Studying Composite Emotions.* Paper presented at the 14th International Conference on Autonomous Agents and Multi-Agent Systems (AAMAS'15).

[56] Lin, J., Yu, H., Shen, Z., & Miao, C. (2014). *Studying Task Allocation Decisions of Novice Agile Teams with Data from Agile Project Management Tools.* Paper presented at the 29th IEEE/ACM International Conference on Automated Software Engineering (ASE'14).

[57] Lin, J., Yu, H., Shen, Z., & Miao, C. (2014). *Using Goal Net to Model User Stories in Agile Software Development.* Paper presented at the 15th IEEE/ACIS International Conference on Software Engineering, Artificial Intelligence, Networking and Parallel/Distributed Computing (SNPD'14).

[58] Liu, S., Miao, C., Liu, Y., Fang, H., Yu, H., Zhang, J., & Leung, C. (2015). *A Reputation Revision Mechanism to Mitigate the Negative Effects of Misreported Ratings.* Paper presented at the 17th International Conference on Electronic Commerce (ICEC'15).

[59] Liu, S., Miao, C., Liu, Y., Yu, H., Zhang, J., & Leung, C. (2015). *An Incentive Mechanism to Elicit Truthful Opinions for Crowdsourced Multiple Choice Consensus Tasks.* Paper presented at the 2015 IEEE/WIC/ACM International Joint Conference on Web Intelligence and Intelligent Agent Technology (WI-IAT'15).

[60]  Liu, S., Yu, H., Miao, C., & Kot, A. C. (2013). *A Fuzzy Logic Based Reputation Model Against Unfair Ratings.* Paper presented at the 12th International Conference on Autonomous Agents and Multi-Agent Systems (AAMAS'13).

[61]  Liu, Y., Liu, S., Fang, H., Zhang, J., Yu, H., & Miao, C. (2014). *RepRev: Mitigating the Negative Effects of Misreported Ratings.* Paper presented at the 28th AAAI Conference on Artificial Intelligence (AAAI-14).

[62]  Liu, Y., Zhang, J., Yu, H., & Miao, C. (2014). *Reputation-aware Continuous Double Auction.* Paper presented at the 28th AAAI Conference on Artificial Intelligence (AAAI-14).

[63]  Mei, J.-P., Yu, H., Liu, Y., Shen, Z., & Miao, C. (2014). *A Social Trust Model Considering Trustees' Influence.* Paper presented at the 17th International Conference on Principles and Practice of Multi-Agent Systems (PRIMA'14).

[64]  Pan, L., Meng, X., Shen, Z., & Yu, H. (2009). *A reputation pattern for service oriented computing.* Paper presented at the Information, Communications and Signal Processing, 2009. ICICS 2009. 7th International Conference on.

[65]  Pan, Z., Miao, C., Tan, B. T. H., Yu, H., & Leung, C. (2015). *Agent Augmented Inter-generational Crowdsourcing.* Paper presented at the 2015 IEEE/WIC/ACM International Joint Conference on Web

Intelligence and Intelligent Agent Technology (WI-IAT'15).

[66] Pan, Z., Miao, C., Yu, H., Leung, C., & Chin, J. J. (2015). *The Effects of Familiarity Design on the Adoption of Wellness Games by the Elderly.* Paper presented at the 2015 IEEE/WIC/ACM International Joint Conference on Web Intelligence and Intelligent Agent Technology (WI-IAT'15).

[67] Pan, Z., Yu, H., Miao, C., & Leung, C. (2016). *Efficient Collaborative Crowdsourcing.* Paper presented at the 30th AAAI Conference on Artificial Intelligence (AAAI-16).

[68] Qin, T., Yu, H., Leung, C., Shen, Z., & Miao, C. (2009). Towards a trust aware cognitive radio architecture. *ACM Sigmobile Mobile Computing and Communications Review, 13*(2), 86-95.

[69] Shen, Z., Yu, H., Miao, C., Li, S., & Chen, Y. (2016). *Multi-Agent System Development MADE Easy.* Paper presented at the 30th AAAI Conference on Artificial Intelligence (AAAI-16).

[70] Shen, Z., Yu, H., Miao, C., & Weng, J. (2011). Trust-based web service selection in virtual communities. *Web Intelligence and Agent Systems, 9*(3), 227-238.

[71] Shi, Y., Sun, C., Li, Q., Cui, L., Yu, H., & Miao, C. (2016). *A Fraud Resilient Medical Insurance Claim System.* Paper presented at the 30th AAAI Conference on Artificial Intelligence (AAAI-16).

[72] Tao, X., Shen, Z., Miao, C., Theng, Y.-L., Miao, Y., & Yu, H. (2011).

Automated negotiation through a cooperative-competitive model *Innovations in Agent-Based Complex Automated Negotiations* (pp. 161-178): Springer Berlin Heidelberg.

[73] Wu, Q., Han, X., Yu, H., Shen, Z., & Miao, C. (2013). *The Innovative Application of Learning Companions in Virtual Singapura.* Paper presented at the Proceedings of the 2013 international conference on Autonomous agents and multi-agent systems.

[74] Yu, H., Cai, Y., Shen, Z., Tao, X., & Miao, C. (2010). *Agents as intelligent user interfaces for the net generation.* Paper presented at the Proceedings of the 15th international conference on Intelligent user interfaces.

[75] Yu, H., Lin, H., Lim, S. F., Lin, J., Shen, Z., & Miao, C. (2015). *Empirical Analysis of Reputation-aware Task Delegation by Humans from a Multi-agent Game.* Paper presented at the 14th International Conference on Autonomous Agents and Multi-Agent Systems (AAMAS'15).

[76] Yu, H., Liu, S., Kot, A. C., Miao, C., & Leung, C. (2011). *Dynamic witness selection for trustworthy distributed cooperative sensing in cognitive radio networks.* Paper presented at the Proceedings of the 13th IEEE International Conference on Communication Technology (ICCT'11).

[77] Yu, H., Miao, C., An, B., Leung, C., & Lesser, V. R. (2013). *A

*Reputation Management Approach for Resource Constrained Trustee Agents.* Paper presented at the 23rd International Joint Conference on Artificial Intelligence (IJCAI'13).

[78] Yu, H., Miao, C., An, B., Shen, Z., & Leung, C. (2014). *Reputation-aware Task Allocation for Human Trustees.* Paper presented at the 13th International Conference on Autonomous Agents and Multi-Agent Systems (AAMAS'14).

[79] Yu, H., Miao, C., Liu, S., Pan, Z., Khalid, N. S. B., Shen, Z., & Leung, C. (2016). *Productive Aging through Intelligent Personalized Crowdsourcing.* Paper presented at the 30th AAAI Conference on Artificial Intelligence (AAAI-16).

[80] Yu, H., Miao, C., & Shen, Z. (2015). Apparatus and Method for Efficient Task Allocation in Crowdsourcing: US Patent App. 14/656,009.

[81] Yu, H., Miao, C., Shen, Z., & Leung, C. (2015). *Quality and Budget aware Task Allocation for Spatial Crowdsourcing.* Paper presented at the 14th International Conference on Autonomous Agents and Multi-Agent Systems (AAMAS'15).

[82] Yu, H., Miao, C., Shen, Z., Leung, C., Chen, Y., & Yang, Q. (2015). *Efficient Task Sub-delegation for Crowdsourcing.* Paper presented at the 29th AAAI Conference on Artificial Intelligence (AAAI-15).

[83] Yu, H., Miao, C., Tao, X., Shen, Z., Cai, Y., Li, B., & Miao, Y. (2009). *Teachable Agents in Virtual Learning Environments: a Case Study.*

Paper presented at the World Conference on E-Learning in Corporate, Government, Healthcare, and Higher Education.

[84] Yu, H., Miao, C., Weng, X., & Leung, C. (2012). *A simple, general and robust trust agent to help elderly select online services.* Paper presented at the Network of Ergonomics Societies Conference (SEANES), 2012 Southeast Asian.

[85] Yu, H., Shen, Z., & An, B. (2012). An Adaptive Witness Selection Method for Reputation-Based Trust Models. *PRIMA 2012: Principles and Practice of Multi-Agent Systems*, 184-198.

[86] Yu, H., Shen, Z., & Leung, C. (2013). Towards Health Care Service Ecosystem Management for the Elderly. *International Journal of Information Technology (IJIT), 19*(2).

[87] Yu, H., Shen, Z., Leung, C., Miao, C., & Lesser, V. R. (2013). A Survey of Multi-agent Trust Management Systems. *IEEE Access, 1*(1), 35-50.

[88] Yu, H., Shen, Z., Li, X., Leung, C., & Miao, C. (2012). Whose Opinions to Trust more, your own or others'? *The 1st Workshop on Incentives and Trust in E-commerce - the 13th ACM Conference on Electronic Commerce (WIT-EC'12)*, 1-12.

[89] Yu, H., Shen, Z., & Miao, C. (2007). *Intelligent software agent design tool using goal net methodology.* Paper presented at the Proceedings of the 2007 IEEE/WIC/ACM International Conference on Intelligent Agent Technology.

[90]  Yu, H., Shen, Z., & Miao, C. (2008). A goal-oriented development tool to automate the incorporation of intelligent agents into interactive digital media applications. *Computers in Entertainment (CIE), 6*(2), 24.

[91]  Yu, H., Shen, Z., & Miao, C. (2009). *A trustworthy beacon-based location tracking model for body area sensor networks in m-health.* Paper presented at the Information, Communications and Signal Processing, 2009. ICICS 2009. 7th International Conference on.

[92]  Yu, H., Shen, Z., Miao, C., & An, B. (2012). *Challenges and Opportunities for Trust Management in Crowdsourcing.* Paper presented at the IEEE/WIC/ACM International Conference on Intelligent Agent Technology (IAT).

[93]  Yu, H., Shen, Z., Miao, C., & An, B. (2013). *A Reputation-aware Decision-making Approach for Improving the Efficiency of Crowdsourcing Systems.* Paper presented at the 12th International Conference on Autonomous Agents and Multi-Agent Systems (AAMAS'13).

[94]  Yu, H., Shen, Z., Miao, C., An, B., & Leung, C. (2014). Filtering Trust Opinions through Reinforcement Learning. *Decision Support Systems (DSS), 66*, 102-113.

[95]  Yu, H., Shen, Z., Miao, C., Leung, C., & Niyato, D. (2010). A survey of trust and reputation management systems in wireless communications. *Proceedings of the IEEE, 98*(10), 1755-1772.

[96] Yu, H., Shen, Z., Miao, C., & Tan, A.-H. (2011). *A simple curious agent to help people be curious.* Paper presented at the 10th International Conference on Autonomous Agents and Multiagent Systems-Volume 3.

[97] Yu, H., Shen, Z., Miao, C., Wen, J., & Yang, Q. (2007). *A service based multi-agent system design tool for modelling integrated manufacturing and service systems.* Paper presented at the Emerging Technologies and Factory Automation, 2007. ETFA. IEEE Conference on.

[98] Yu, H., & Tian, Y. (2005). *Developing Multiplayer Mobile Game Using MIDP 2.0 Game API and JSR-82 Java Bluetooth API.* Paper presented at the 2005 Cybergames Conference.

[99] Yu, H., Yu, X., Lim, S. F., Lin, J., Shen, Z., & Miao, C. (2014). *A Multi-Agent Game for Studying Human Decision-making.* Paper presented at the 13th International Conference on Autonomous Agents and Multi-Agent Systems (AAMAS'14).

[100] Z Man, K Lee, D Wang, Z Cao, and C Miao, "A new robust training algorithm for a class of single-hidden layer feedforward neural networks," Neurocomputing 74 (16), 2491-2501, 2011.

[101] L Pan, X Luo, X Meng, C Miao, M He, and X Guo, "A Two-Stage Win–Win Multiattribute Negotiation Model: Optimization and Then Concession," Computational Intelligence 29 (4), 577-626, 2013.

[102] YZ Zhao, M Ma, CY Miao, and TN Nguyen, "An energy-efficient and low-latency MAC protocol with Adaptive Scheduling for multi-hop

wireless sensor networks," Computer Communications 33 (12), 1452-1461, 2010.

[103] J Weng, C Miao, A Goh, Z Shen, and R Gay, "Trust-based agent community for collaborative recommendation," In Proceedings of the 5th international joint conference on Autonomous agents and multi-agent systems (AAMAS'06), 1260-1262, 2006

[104] HJ Song, CY Miao, ZQ Shen, W Roel, DH Maja, C Francky, "Design of fuzzy cognitive maps using neural networks for predicting chaotic time series," Neural Networks 23 (10), 1264-1275, 2010.

[105] HJ Song, ZQ Shen, CY Miao, Y Miao, and BS Lee, "A fuzzy neural network with fuzzy impact grades," Neurocomputing 72 (13), 3098-3122, 2009.

[106] C Miao, Q Yang, H Fang, and A Goh, "Fuzzy cognitive agents for personalized recommendation Web Information Systems Engineering,", In *Proceedings of the 3rd International Conference on Web Information Systems Engineering (WISE'02)*, 362-371, 2002.

[107] YZ Zhao, C Miao, M Ma, JB Zhang, and C Leung, "A survey and projection on medium access control protocols for wireless sensor networks," ACM Computing Surveys (CSUR) 45 (1), 7, 2012.

[108] DS Domazet, MC Yan, CFY Calvin, HPH Kong, and A Goh, "An infrastructure for inter-organizational collaborative product development System Sciences," In *Proceedings of the 33rd Annual*

*Hawaii International Conference on System Sciences*, 2000.

[109] G Zhao, Z Shen, C Miao, and Z Man, "On improving the conditioning of extreme learning machine: a linear case," In *Proceedings of the 7th International Conference on Information, Communications and Signal Processing (ICICS'09)*, 2009.

[110] C Miao, A Goh, Y Miao, and Z Yang, "A dynamic inference model for intelligent agents," International Journal of Software Engineering and Knowledge Engineering 11 (05), 509-528, 2001.